\documentclass[twocolumn,prb,showpacs,preprintnumbers,amsmath,amssymb,citeautoscript]{revtex4-2}
\usepackage{graphicx}
\usepackage{dcolumn}
\usepackage{bm}
\usepackage{epstopdf}

\usepackage{color,hyperref}
\definecolor{blue}{rgb}{0.0,0.0,1}
\hypersetup{colorlinks,breaklinks,
            linkcolor=blue,urlcolor=blue,
            anchorcolor=blue,citecolor=blue}

\begin{document}
\preprint{MC-MET-CGVdW}

\title{Native defects and erbium impurities in CaWO$_4$}

\author{Minseok Choi$^{1, 2, 3}$}
\email{minseok.choi@inha.ac.kr} 
\author{Mark E. Turiansky$^{3, 4}$}
\author{BaiQing Zhao$^{3}$}
\author{Jeff D. Thompson$^{5}$}
\author{Chris G. Van de Walle$^{3}$}
\email{vandewalle@mrl.ucsb.edu}
\affiliation{$^{1}$Department of Physics, Inha University, Incheon 22212, Korea}
\affiliation{$^{2}$Institute of Quantum Science, Inha University, Incheon 22212, Korea}
\affiliation{$^{3}$Materials Department, University of California, Santa Barbara, CA 93106-5050, USA}
\affiliation{$^{4}$US Naval Research Laboratory, 4555 Overlook Avenue SW, Washington, DC 20375, USA}
\affiliation{$^{5}$Department of Electrical Engineering, Princeton University, Princeton, NJ 08544, USA}

\date{\today}

\begin{abstract}

We perform hybrid density functional calculation to study the energetics, electronic properties, optical transitions, and migration barriers of native defects in CaWO$_4$. 
Oxygen and calcium vacancies are most likely to form in the absence of doping, but interstitials could also incorporate.
%Oxygen-related defects [O vacancies ($V_{\rm O}$) and O interstitials (O$_i$)] and calcium-related defects [Ca vacancies ($V_{\rm Ca}$) and Ca interstitials (Ca$_i$)] are most likely to form. 
Tungsten-related defects are unlikely to be present. 
The positively charged $V_{\rm O}$ and the negatively charged $V_{\rm Ca}$ are likely to form complexes.
Calculated optical transition levels indicate that experimentally observed absorption and emission peaks can be attributed mainly to oxygen-related defects. 
%Chris 051326
Calculations of migration barriers allow us to conclude that Ca$_i^{2+}$, $V_{\rm O}^{2+}$ and O$_i^{2-}$ are highly mobile, even below room temperature.
We have also examined Er dopants, finding that erbium easily substitutes on the Ca site in a positive charge state. 
Erbium can form complexes with $V_{\rm Ca}$ and O$_i$, which would deactivate the Er. 
%Chris 051326
If Er is introduced by implantation, Er interstitials are likely present, which will produce emission that is prone to spectral diffusion and blinking.  
Our calculated properties of Er$_i$ explain why annealing at modest temperatures allows the interstitials to move into substitutional sites and point defects to move away, resulting in stable emission.

\end{abstract}

%\keywords{Suggested keywords}

\maketitle

\section{Introduction}

\textcolor{red}{}

\begin{figure}[]
\includegraphics[width = 6 cm]{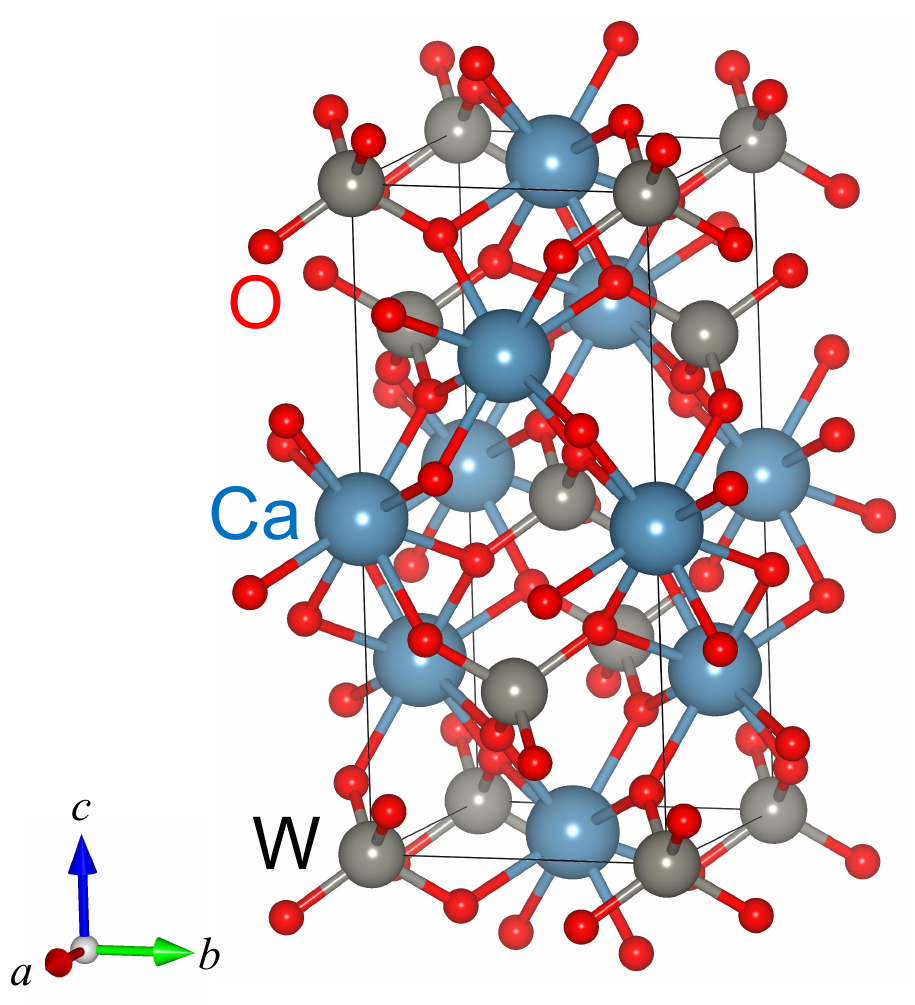}\\
\caption{\label{fig_crystal} Crystal structure of scheelite CaWO$_4$. The conventional unit cell containing 24 atoms is shown. All the lattice sites for each atomic species are symmetry-equivalent. Each Ca atom is bonded to eight O atoms, and each W to four O. Each O is surrounded by one W atom and two Ca atoms.  
}
\end{figure}

CaWO$_4$ crystallizes in the scheelite structure (Fig.~\ref{fig_crystal}), which is tetragonal ($I4_1/a$ space group). 
The material has attracted attention due to its high chemical stability, high photo-sensitivity, scintillating properties, and wide visible emission spectra~\cite{03_Pang_JPCM,2020_CWO_ER,Eg_B3LYP_2022,Lanfranchi2013}. 
Its optical properties can be engineered by adding select rare-earth ions. 
%Eu$^{3+}$, Dy$^{3+}$, Sm$^{3+}$, and Er$^{3+}$ doping into CaWO$_4$ films under ultraviolet irradiation results in the emission of red, yellow, orange, and green light with comparably high quantum yields, respectively~\cite{03_Pang_JPCM}. 
Doped CaWO$_4$ films emit visible light with high quantum yields under ultraviolet irradiation, with Eu(III) producing red, Dy(III) yellow, Sm(III) orange, and Er(III) green emission~\cite{03_Pang_JPCM} (where the roman numerals indicate oxidation states). 
Tb-doped powders prepared by a solid-state reaction emit green light~\cite{2023_Tb_CWO}. 
%In the absence of doping, 
The optical properties of undoped CaWO$_4$ are known to be very sensitive to the synthesis and deposition techniques, implying that native defects 
%such as oxygen vacancies ($V_{\rm O}$) 
play a role. 
%In general, adequate knowledge about CWO is lacking; even the value of the fundamental band gap is still unclear, with reported values in the range of 5 to 6 eV~\cite{2012_PL_CWO_Eg,2004_Mikhailik_PRB_Eg,2006_CWO_Eg,2002_Kolobanov_Eg}.
%Tb Eru3 doping into CaWO$_4$ nanopartilces under ultraviolet irradiation result in the emission of green and red light with comparably high quantum yields, respectively~\cite{2012_Mai_CWO_REI,2018_CWO_Eu}. 

CaWO$_4$ has been flagged as an excellent candidate for hosting rare-earth ions, providing highly coherent 4$f$-4$f$ optical and spin transitions, thus serving as a materials platform for quantum communication and information technologies~\cite{2023_REVIEW_REI_QauntumNetwork}.
%In addition, it has been predicted that the natural isotopic distribution of CaWO$_4$ results in minimal nuclear spin noise and enables spin qubits with long quantum coherence times~\cite{2021_Bertet_SciAdv_ErCWO}.
%Er embedded in CaWO$_4$ crystals has been found to emit photons in the telecom-band wavelength (1.5 $\mu$m) with a long coherence time (23 ms)~\cite{2021_Bertet_SciAdv_ErCWO} for the Er$^{3+}$ electron spin; it thus shows promise for use in quantum repeaters to boost signals between network nodes for long-distance quantum networking~\cite{2021_Bertet_SciAdv_ErCWO,2023_Nature_Jeff_ErCWO}. 
Er embedded in CaWO$_4$ crystals has been found to emit photons in the telecom-band wavelength (1.5 $\mu$m) with a long coherence time (23 ms)~\cite{2021_Bertet_SciAdv_ErCWO} for the Er$^{3+}$ electron spin.
Indeed, the natural isotopic distribution of CaWO$_4$ results in minimal nuclear spin noise and enables spin qubits with long quantum coherence times~\cite{2022_Galli_PNAS}.
% ($T_2$ = 18 ms)
Er in CaWO$_4$ thus shows promise for use in quantum repeaters to boost signals between network nodes for long-distance quantum networking~\cite{2021_Bertet_SciAdv_ErCWO,2023_Nature_Jeff_ErCWO}.
However, defects and impurities affect spin, charge, and optical properties, resulting in decoherence and noise~\cite{2023_Nature_Jeff_ErCWO,2021_Wolfowicz_NRP,2022_PRB_Jeff_Er_SR}. 
%Shin {\it et al.}~\cite{2022_Shin_ErTiO2} measured an additional photoluminescence (PL) peak in Er-doped TiO$_2$ films, which was attributed to oxygen vacancies.
%$V_{\rm O}$ formation caused by charge imbalance introduced by Er dopants. 
%In Er-implanted oxides such as PbWO$_4$ and MgO, an increase in inhomogeneous linewidth was observed in photoluminescence(PL)-excitation spectroscopy and associated with implantation-induced disorder~\cite{2022_PRB_Jeff_Er_SR}.

% However, the amount of decoherence experienced by an rare-earth ion in materials is determined by the amount of noise (arising from other spins, charges and lattice vibrations in the host crystal) and the sensitivity of the defect to that noise~\cite{2023_Nature_Jeff_ErCWO_2}
 
%CWO provides some benefits for the applications: Er$^{3+}$ ions are incorporated at a non-polar site, reducing the sensitivity to charge, and the low abundance of nuclear spins, reducing magnetic noise, allows for a long spin-coherence time~\cite{2023_Nature_Jeff_ErCWO,2023_Nature_Jeff_ErCWO_2}

A few first-principles studies of defects in CaWO$_4$ have been reported~\cite{Zhang_PBE_Vo_2008,2008_DVXA_VO,2008_defect_GULP,Ackerman_Vcation_VASP_2016,2023_CWO_PBEU}. 
%They later used the discrete variational-X$\alpha$ method and found that $V_{\rm O}^+$ can be responsible for the peak at 520 nm~\cite{2008_DVXA_VO}.
Using density functional theory (DFT) with the generalized gradient approximation (GGA), Shao {\it et al.}~\cite{Zhang_PBE_Vo_2008} showed that oxygen vacancies ($V_{\rm O}$) can introduce states near mid-gap and suggested that $V_{\rm O}$ is responsible for the experimentally observed absorption bands at 340 nm (3.65 eV) and 520 nm (2.38 eV)~\cite{Baccaro_pssa_2000}.
Ackerman {\it et al.} performed DFT-GGA calculations for defect formation energies of cation vacancies as well as Te impurities~\cite{Ackerman_Vcation_VASP_2016}, and found that the Ca vacancy ($V_{\rm Ca}$) has a lower formation energy than the W vacancy ($V_{\rm W}$). 
Another computational study by Shao {\it et al.}~\cite{2008_defect_GULP} reported defect energetics and structure for several native defects using an interatomic potential scheme. 
The authors suggested that the predominant intrinsic defects should be Frenkel-type defects on the oxygen site,
and that W interstitials (W$_i$) are not energetically stable. 
%It was also suggested that W interstitials (W$_i$) would not exist in CaWO$_4$ (the configuration energetically diverges). 
More recently, Ferri {\it et al.}~\cite{2023_CWO_PBEU} performed GGA+$U$ calculations (with $U$ = 3~eV for W 5$d$) and analyzed formation energies of native defects by considering various atomic chemical potentials, mimicking experimental growth environments. 
They found that $V_{\rm Ca}$ and O interstitials (O$_i$) are predominant under O-rich and intermediate conditions, whereas $V_{\rm O}$ is most likely to form under W-rich conditions. 

%None of these DFT studies used state-of-the-art hybrid functionals, and a detailed understanding of the microscopic structure and energetics and their dependence on charge states and atomic chemical potentials is still lacking. 
In spite of these DFT studies, a detailed understanding of the microscopic structure and energetics and their dependence on charge states and atomic chemical potentials is still lacking. 
In the present work, we use the Heyd-Scuseria-Ernzerhof (HSE) hybrid density functional~\cite{heyd:8207,krukau:224106}, which has been demonstrated to result in accurate band structures and enthalpies of formation, and has provided a reliable description of defect formation energies and defect levels in wide-band-gap materials~\cite{2010_AGali_PRB,RMP_2014,2016_AA_tutorial,2017_Hinuma_PRB_BandOffset,2018_oba_review}.   
%In addition, most of these studies were carried out using the DFT-GGA, in which the band gap is severely underestimated, leading to errors in defect formation energies and in the position of transition levels with respect to the band edges. 
We perform a systematic study of native defects, addressing their electronic properties and optical transition.
We also calculate migration barriers, and examine the properties of Er impurities and their interaction with native defects.

%Computational details of the DFT and defect formation energy are provided in Sec.~\ref{sec:method}.  The defect formation energies under various growth conditions are addressed in Sec.~\ref{sec:natdef}, and the electronic and structural properties of each defect are described in Sec.~\ref{sec:natdef}.

%%%%%%%%%%%%%%%%%%%%%%%%%%%%%%%%%%%%%%%%%%%%%%%%%%%%%%
%%%%%%%%%%%%%%%%%%%%%%%%%%%%%%%%%%%%%%%%%%%%%%%%%%%%%%
\section{Computational approach}
\label{sec:method}

\subsection{Hybrid density functional calculations}
\label{sec:dft}

First-principles DFT calculations were carried out using the HSE hybrid functional,~\cite{heyd:8207,krukau:224106} implemented with the projector augmented-wave method \cite{paw} in the {\sc vasp} code \cite{vasp}. The HSE mixing parameter was set to 25\%. The electronic wave functions were expanded in a plane-wave basis set with an energy cutoff of 500 eV. The pseudopotentials were used in which 3$p^6$4$s^2$ for Ca, 6$s^1$5$d^5$ for W, 2$s^2$2$p^4$ for O, and 5$p^6$5$d^1$6$s^2$ for Er were treated as valence electrons.

For bulk calculations, we used the conventional unit cell of scheelite CaWO$_4$ which is tetragonal and contains 24 atoms (Fig.~\ref{fig_crystal}). The integrations over the Brillouin zone of the bulk cell were performed using a 4$\times$4$\times$2 $k$-point grid, and  atomic positions were relaxed until the Hellmann-Feynman forces were less than 0.01 eV/\AA.
Our HSE calculations produce results for the structure in good agreement with experiment, as can be seen in Table~\ref{param}.
Our calculated band-gap energy of 5.62 eV is significantly larger than the values obtained in GGA (3.5 eV)~\cite{Zhang_PBE_Vo_2008} or GGA+$U$ (4.15 eV)~\cite{2023_CWO_PBEU} calculations. 
Our value lies within the range of experimentally reported values~\cite{2004_Mikhailik_PRB_Eg,2006_CWO_Eg,2002_Kolobanov_Eg} (see Table~\ref{param})  and is similar to the value of 5.71 eV obtained with the B3LYP hybrid functional~\cite{Eg_B3LYP_2022}.
%Our value lies within the range of experimentally reported values~\cite{2012_PL_CWO_Eg,2004_Mikhailik_PRB_Eg,2006_CWO_Eg,2002_Kolobanov_Eg} (see Table~\ref{param})  and is similar to the value of 5.71 eV obtained with the B3LYP hybrid functional~\cite{Eg_B3LYP_2022}.

\begin{table}
\caption{\label{param} Lattice parameters and band gap of scheelite CaWO$_4$ from HSE calculations and experiment.  }
\begin{ruledtabular}
\begin{tabular}{cccc}
%&  \multicolumn{3}{c}{x}  \\
%\cline{2-4}
     Property  &  HSE &Experiment \\
    \hline
    $a$ (\AA)  &   5.260 & 5.243~\cite{1964_CWO_xrd}\\
    $c$ (\AA)  &    11.413 & 11.376~\cite{1964_CWO_xrd}\\
        $d_{\rm Ca - O}$ (\AA)  &   2.451, 2.485 & 2.44, 2.48~\cite{1964_CWO_xrd}\\
                $d_{\rm W - O}$ (\AA)  & 1.787 &1.78~\cite{1964_CWO_xrd}\\
    $E_{\rm g}$ (eV) & 5.618 & 5.2~\cite{2004_Mikhailik_PRB_Eg} 5.4~\cite{2006_CWO_Eg} 6.0~\cite{2002_Kolobanov_Eg}\\
%    $E_{\rm g}$ (eV) & 5.618 & 4.9$\pm$0.15 eV~\cite{2012_PL_CWO_Eg},  5.2$\pm$0.3~\cite{2004_Mikhailik_PRB_Eg} 5.4~\cite{2006_CWO_Eg}, $\sim$6~\cite{2002_Kolobanov_Eg}\\
\hline
\end{tabular}
\end{ruledtabular}
\end{table}

Defect calculations were performed using 96-atom supercells constructed by taking a $2 \times 2 \times 1$ multiple of the conventional unit cell.  
Brillouin-zone integrations used a 2$\times$2$\times$2 $k$-point grid, and atomic positions were relaxed until the Hellmann-Feynman forces were less than 0.02 eV/\AA. 
Spin polarization was included. 
We tested convergence as a function of supercell size (going up to 324-atom cells, which are a $3 \times 3 \times 3$  multiple of the 12-atom primitive cell) for neutral $V_{\rm O}$ and found that Kohn-Sham states changed by less than 0.02~eV, and bond lengths by less than 1\%. 
%the distances between $V_{\rm O}$ and the nearest cations of 

%With Eb=0.91 eV for Hi , we obtain a temperature of 328 K, i.e., hydrogen will become mobile at temperatures slightly above room temperature.

%%%%%%%%%%%%%%%%%%%%%%%%%%%%%%%%%%%%%%%%%%%%%%%%%%%%%%
\subsection{Formation energies and transition levels}
%%%%%%%%%%%%%%%%%%%%%%%%%%%%%%%%%%%%%%%%%%%%%%%%%%%%%%

The formation energy of a defect or impurity $X$ in charge state $q$ is given by~\cite{RMP_2014}:
\begin{eqnarray} \label{eq:form}
\  E^{f} (X^q) = E_{\rm tot}(X^q) - E_{\rm tot}({\rm CaWO_4})\nonumber\\
- \sum_{i} n_i (\mu_i^0 + \mu_i) + q\epsilon_{F} + \Delta^q,
\end{eqnarray}
where $E_{\rm tot}(X^q)$ is the total energy of a supercell containing the defect or dopant $X$ in charge state $q$, and $E_{\rm tot}({\rm CaWO_4})$ is the total energy of the pristine CaWO$_4$ in the same supercell. $n_i$ is the number of atoms of type $i$ added ($n_i$$>$0) to or removed from ($n_i$$<$0) the pristine CaWO$_4$, and $\mu_{i}$ are chemical potentials that reflect the abundance of elements in a growth or processing environment, as discussed below.  
$\epsilon_{F}$ is the Fermi energy, which will be referenced to the valence-band maximum (VBM).
$\Delta^q$ is a term that corrects for interactions between supercells and with compensating background charges.~\cite{Freysoldt_09,Freysoldt_11}

The charge-state transition level $(q/q')$ is defined as the Fermi-level position below which the defect is most stable in charge state $q$ and above which the same defect is stable in charge state $q'$:
\begin{equation} \label{level}
(q/q') = \frac{E^f (X^q;\epsilon_{F}=0) - E^f (X^{q'};\epsilon_{F}=0)}{(q'-q)},
\end{equation}
where $E^f(X^q;\epsilon_{F}=0)$ is the formation energy for $X^q$ when $\epsilon_{F}$ is at the VBM. The charge-state transition levels are not affected by the choice of chemical potentials.
Optical transitions are calculated based on formation energies and using the Frank-Condon principle, i.e., assuming vertical transitions in which the atomic configuration is kept fixed when changing the charge state of the defect, thus providing values for the peaks in absorption and emission spectra~\cite{RMP_2014}.

%%%%%%%%%%%%%%%%%%%%%%%%%%%%%%%%%%%%%%%%%%%%%%%%%%%%%%
\subsection{Atomic chemical potentials}
%%%%%%%%%%%%%%%%%%%%%%%%%%%%%%%%%%%%%%%%%%%%%%%%%%%%%%

The chemical potentials are variables that reflect the relative abundance of the constituents during growth or processing. 
The atomic chemical potential $\mu_{i}$ is referenced to the total energy per atom of the standard phase of the species $\mu_{i}^0$:
$\mu_{\rm Ca}^0$ is the total energy per atom of Ca metal, $\mu_{\rm W}^0$ that of W metal, $\mu_{\rm Er}^0$ that of Er metal, and $\mu_{\rm O}^0$ is half the total energy of an isolated O$_2$ molecule. 
%For $\mu_{\rm Er}^0$ we use Er metal.

\begin{figure}[h]
\includegraphics[width = 6 cm]{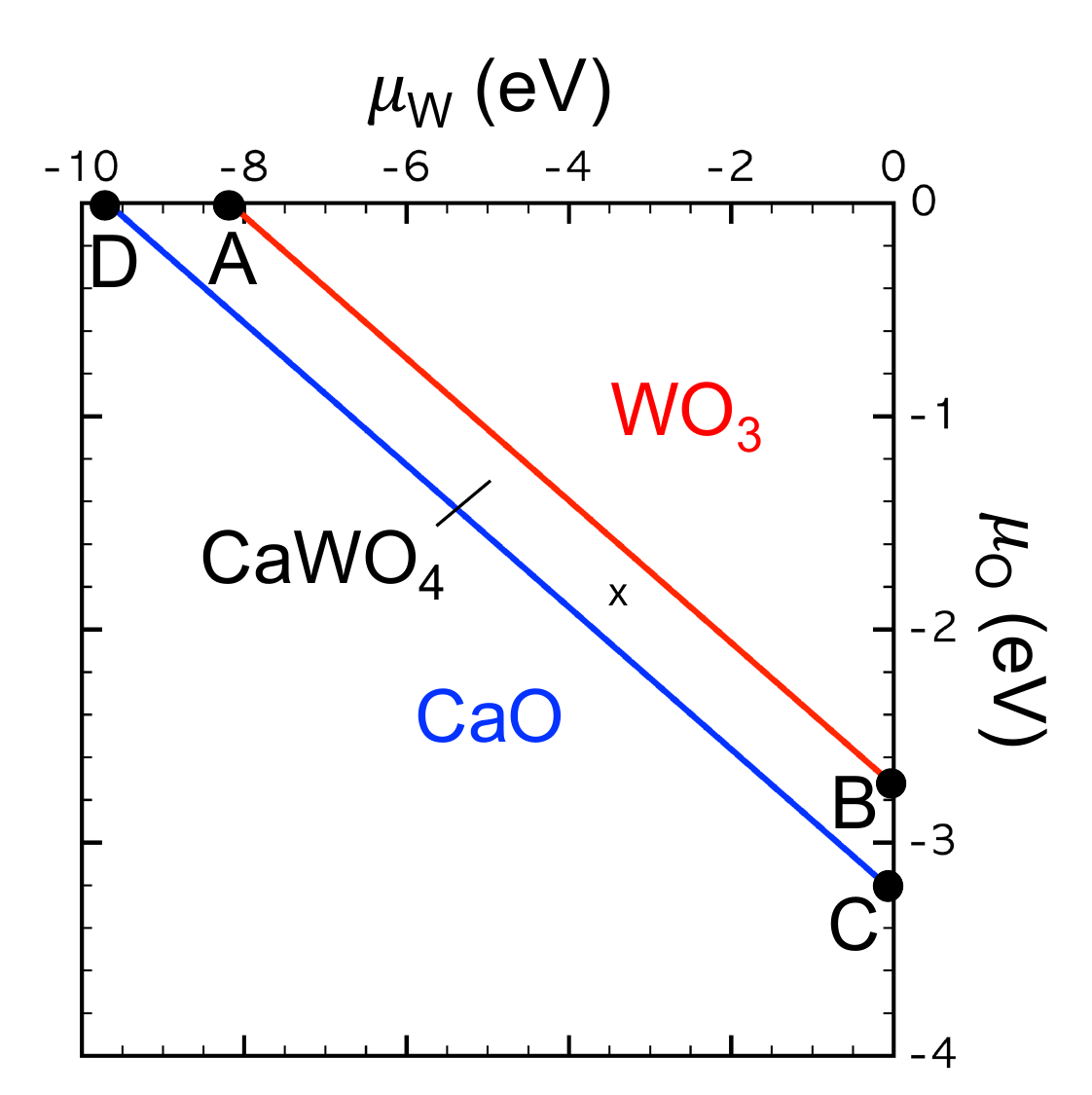}\\
\caption{\label{chempot} Phase stability diagram of CaWO$_4$, showing the region in the chemical potential plane in which CaWO$_4$ is stable. The chemical potentials $\mu_\mathrm{Ca}$, $\mu_\mathrm{W}$, and $\mu_\mathrm{O}$ are limited by the formation of the secondary phases as described in the text and Table. 
Chemical-potential conditions corresponding to representative film growth conditions, as discussed in Sec.~\ref{sec:prev}, are indicated by the ``x'' symbol.
%CWO is not stable at Ca-rich limit ($\mu_\mathrm{Ca}$ = 0).
}
\end{figure}

The chemical potentials $\mu_{i}$ ($i$ = Ca, W, and O) are restricted by the formation of limiting phases. They must satisfy the stability condition of CaWO$_4$:
\begin{equation} \label{Eq3}
\  \mu_{\rm Ca} +  \mu_{\rm W}+ 4 \mu_{\rm O} = \Delta H _f {\rm (CaWO_4)},
\end{equation}
where $\Delta H_f {\rm (CaWO_4)}$ is the formation enthalpy of the CaWO$_4$ crystal.
To prevent the formation of the elemental phases we must have $\mu_{\rm Ca}$ $\le$ 0, $\mu_{\rm W}$ $\le$ 0, and $\mu_{\rm O}$ $\le$ 0, and 
the chemical potentials are further constrained by the formation of CaO and WO$_3$ phases:
\begin{eqnarray} \label{Eq4}
\ \mu_{\rm Ca} + \mu_{\rm O} \le \Delta H _f {\rm (CaO)},
\end{eqnarray}
\begin{eqnarray} \label{Eq5}
\ \mu_{\rm W} + 3\mu_{\rm O} \le \Delta H _f {\rm (WO_3)},
\end{eqnarray}
where $\Delta H _f {\rm (CaO)}$ and $\Delta H _f {\rm (WO_3)}$ are the formation enthalpies of CaO and WO$_3$. 
Our calculated formation enthalpies (compared with experimental values in parentheses) are --6.09 eV (--6.58~eV~\cite{chase1998nist}) for CaO, --8.17 eV (--8.73~eV~\cite{1959_WO3_Mah}) for WO$_3$, and --15.77~eV (--17.06~eV~\cite{1996_CWO_guo}) for CaWO$_4$.  
We use Eqs.~(\ref{Eq3})--(\ref{Eq5}) to define a region in the chemical potential plane in which CaWO$_4$ is thermodynamically stable.  This region is illustrated in Fig.~\ref{chempot}.
To focus our discussion we select specific sets of chemical-potential values, as shown in Fig.~\ref{chempot} and listed in Table~\ref{enthalpy}. 
For the Er chemical potential, Er$_2$O$_3$ was used as a limiting phase, and its calculated (experimental) formation enthalpy is --19.14 eV (--19.74 eV~\cite{MORSS1993415}).

\begin{table}
\caption{\label{enthalpy} Values of atomic chemical potential $\mu_i$ ($i$ = Ca, W, and O) at the conditions indicated in Figure~\ref{chempot}. The values are obtained based on calculated formation enthalpies for CaO, WO$_3$, and CaWO$_4$.}
\begin{ruledtabular}
\begin{tabular}{cccc}
     Condition  & $\mu_\mathrm{Ca}$ (eV)  &   $\mu_\mathrm{W}$ (eV)  &  $\mu_\mathrm{O}$ (eV)\\
\hline
Point A &  --7.59 & --8.18  &  0\\
Point B &  --4.85 & 0   &  --2.73\\
Point C & --2.85 & 0   &  --3.23\\
Point D & --6.09 &--9.68   &  0\\
\end{tabular}
\end{ruledtabular}
\end{table}
%%%%%%%%%%%%%%%%%%%%%%%%%%%%%%%%%%%%%%%%%%%%%%%%%%%%%%
%%%%%%%%%%%%%%%%%%%%%%%%%%%%%%%%%%%%%%%%%%%%%%%%%%%%%%
\subsection{Migration barriers}
%%%%%%%%%%%%%%%%%%%%%%%%%%%%%%%%%%%%%%%%%%%%%%%%%%%%%%
\label{ssec:migration}

Migration barriers ($E_{\rm b}$) of several important defects were calculated using the climbing-image nudged elastic band (cNEB) method~\cite{2000_cNEB}. %To mitigate computational cost, we used the GGA functional~\cite{pbe} in the cNEB calculations based on the initial and barrier geometries of the HSE total energy calculations. We tested the accuracy of this approach for the migration of O vacancy; full HSE calculations yield migration barriers that are within 0.1 eV of the results obtained using PBE calculations.

The migration barrier allows us to estimate the temperature at which the impurity becomes mobile. Within transition state theory~\cite{VINEYARD1957121}, the rate $\Gamma$ at which an impurity hops to a neighboring equivalent site can be expressed as
\begin{eqnarray} \label{eq:migration}
\  {\Gamma} = {\Gamma_0} \, {\rm exp} (\frac{-E_{\rm b}}{k_{\rm B}T}),
\end{eqnarray}
where $k_{\rm B}$ is the Boltzmann constant. The prefactor $\Gamma_0$, related to a typical vibrational frequency, can be approximated as $10^{14}$ s$^{-1}$. An estimate for the annealing temperature $T_a$ at which the impurity becomes mobile can then be obtained as the temperature at which the rate $\Gamma$=1 s$^{-1}$, which leads to $T_a$ $\approx$ $E_{\rm b}$ $\times$ 360 K/eV. 

%%%%%%%%%%%%%%%%%%%%%%%%%%%%%%%%%%%%%%%%%%%%%%%%%%%%%%

\section{Results and discussion}
\label{sec:natdef}

%%%%%%%%%%%%%%%%%%%%%%%%%%%%%%%%%%%%%%%%%%%%%%%%%%%%%%
%\subsection{Defect formation energies}
\subsection{Native defects}
%%%%%%%%%%%%%%%%%%%%%%%%%%%%%%%%%%%%%%%%%%%%%%%%%%%%%%
\label{sec:form}

%The Ca vacancy ($V_{\rm Ca}$), the W vacancy ($V_{\rm W}$), the O vacancy ($V_{\rm O}$), the Ca interstitial (Ca$_i$), the W interstitial (W$_i$), and the O interstitial (O$_i$) were considered.

\begin{figure*}[]
\includegraphics[width = 16 cm]{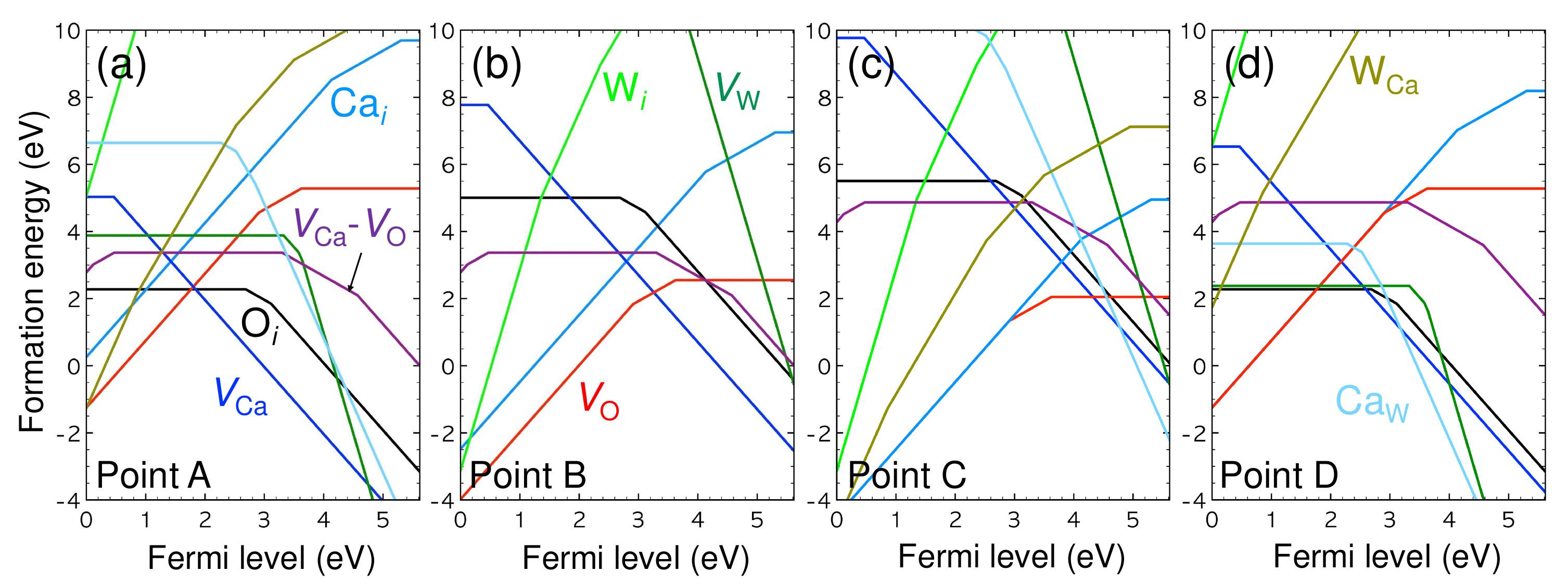}\\
\caption{ \label{fig:fig_formE}  Formation energies of native point defects as a function of the Fermi level at four representative conditions as specified in Fig.~\ref{chempot} and Table~\ref{enthalpy}: (a) O-rich, W-rich; (b) O-poor, W-rich; (c) O-poor, Ca-rich; (d) O-rich, Ca-rich.}
\end{figure*}

In Fig.~\ref{fig:fig_formE}, we show the defect formation energies as a function of Fermi-level position in the band gap for the four chemical-potential conditions considered in Fig,~\ref{chempot} and Table~\ref{enthalpy}: points A and D under O-rich conditions and points B and C under W-rich conditions. 
Defect formation energies at the other points in the region where CaWO$_4$ is stable can be estimated using Eq.~(\ref{eq:form}), Fig.~\ref{fig:fig_formE}, and Table~\ref{enthalpy}. 
The slopes of the formation-energy lines indicate the charge state of the defect, and the kinks in the lines correspond to the position of the charge-state transition levels in the band gap.  

%%%%%%%%%%%%%%%%%%%%%%%%%%%%%%%%%%%%%%%%%%%%%%%%%%%%%%

%%%%%%%%%%%%%%%%%%%%%%%%%%%%%%%%%%%%%%%%%%%%%%%%%%%%%%

In the following sub sections we will discuss details for individual defects, and in Sec.~\ref{sec:prev} we will comment on which native defects are expected to prevail, based on the likely position of the Fermi level.

%%%%%%%%%%%%%%%%%%%%%%%%%%%%%%%%%%%%%%%%%%%%%%%%%%%%%%
\subsubsection{Oxygen vacancies}
%%%%%%%%%%%%%%%%%%%%%%%%%%%%%%%%%%%%%%%%%%%%%%%%%%%%%%
\label{sec:VO}

Each O atom bonds to one W and two Ca atoms. 
In the neutral charge state, $V_{\rm O}^{0}$, the Ca atoms are displaced slightly outward by 0.04 {\AA} and 0.06 {\AA}, and the W atom by 0.04 {\AA}.
In the + charge state, $V_{\rm O}^{+}$, the Ca atoms relax outward by 0.16 {\AA} and 0.18 {\AA}, and the W by 0.06 {\AA}.
In the case of $V_{\rm O}^{2+}$, finally, the Ca are displaced outward by 0.25 {\AA} and 0.26 {\AA} and the W by 0.10 {\AA}.

\begin{figure}[t]
\includegraphics[width = 9  cm]{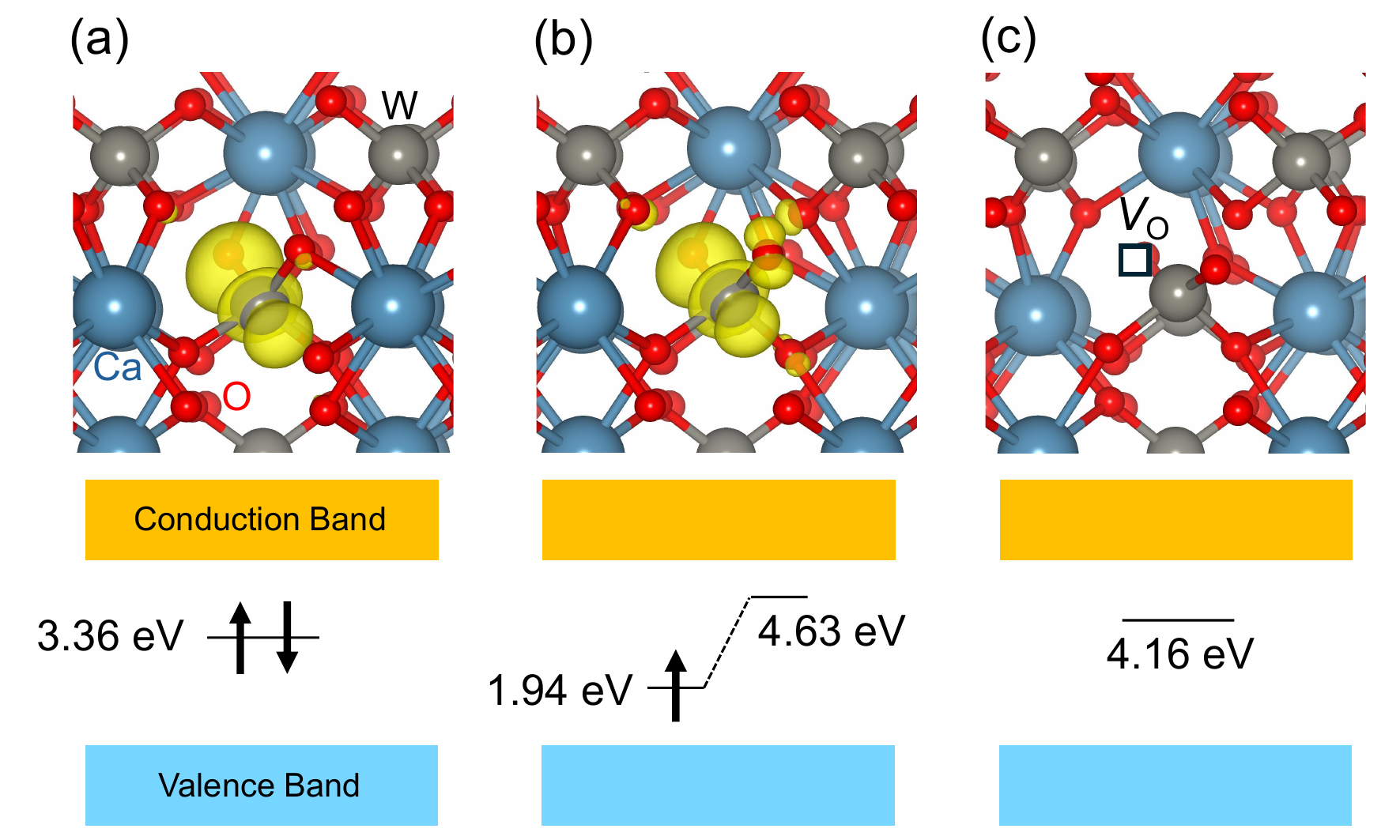}\\
\caption{\label{VO} Local atomic structure for O vacancies in CaWO$_4$ in (a) the neutral ($V_{\rm O}^{0}$), (b) the +  ($V_{\rm O}^{+}$), and (c) the 2+ ($V_{\rm O}^{2+}$) charge state. The position of the oxygen vacancy is indicated by a square in panel (c).
The corresponding Kohn-Sham states in the band gap, along with their occupation, are shown in the lower panels. The charge densities of the occupied gap state for $V_{\rm O}^{0}$ and $V_{\rm O}^{+}$ are also shown.  The isosurfaces correspond to 10\% of the maximum~\cite{VESTA_11}.
}
\end{figure}

The charge density of the defect states in the band gap and the local atomic relaxations are shown in Fig.~\ref{VO}. 
The in-gap Kohn-Sham states are mainly characterized by the 5$d$ orbital of the neighboring W atom and are positioned at 3.36 eV above the VBM for $V_{\rm O}^{0}$, 1.94 eV  and 4.63 eV for $V_{\rm O}^{+}$,  and 4.16 eV for $V_{\rm O}^{2+}$. 
$V_{\rm O}^+$ has 
%a high-spin ground state configuration of 
spin 1/2, i.e., an unpaired spin that could contribute to magnetic noise~\cite{chatterjee_semiconductor_2021,onizhuk_decoherence_2024} and limit the coherence time of spin qubits.

%These deep in-gap states are consistent with previous study (Ref. ~\onlinecite{Zhang_PBE_Vo_2008}) reporting the Kohn-Sham states of $V_{\rm O}^{+2}$ at the mid-gap region. 

As shown in Fig.~\ref{fig:fig_formE}, O vacancies introduce two charge-state transition levels in the upper part of the band gap:
a (2+/+) level at 2.71 eV below the conduction-band minimum (CBM) and a (+/0) level at 1.99 eV below the CBM. 
$V_{\rm O}$ thus acts as a deep donor, in agreement with a recent GGA+$U$ report~\cite{2023_CWO_PBEU}. %We calculated the $V_{\rm O}$-associated vertical optical transitions based on the Franck-Condon principles. The transition $V_{\rm O}^{+} \rightarrow V_{\rm O}^{2+} + e^-$ give rises to an absorption peak at 3.82 eV , and  $V_{\rm O}^{0} \rightarrow V_{\rm O}^{+} + e^-$ at 2.70 eV. These values are close to the experimentally observed at 340 nm (3.65 eV) and 520 nm (2.38 eV) absorption bands~\cite{Zhang_PBE_Vo_2008}. The transition $V_{\rm O}^{+} \rightarrow V_{\rm O}^{2+} + e^-$

Using the cNEB method we calculated the migration barrier for $V_{\rm O}^{2+}$, which is a likely charge state since the CaWO$_4$ samples are insulating and hence the Fermi level is likely to be near mid-gap.
We obtained a barrier $E_{\rm b}$ of 0.43 eV;
%which is comparable to the interstitial hydrogen in oxides~\cite{2022_Sai_AlGaO_PRB}. 
the low value 
%can be attributed to a lack of covalent bonding in the W--O bonds, and 
indicates that the defect is highly mobile (even below room temperature) since  the corresponding $T_a$ is 155 K.
Oxygen vacancies are thus unlikely to be present as isolated defects.

%%%%%%%%%%%%%%%%%%%%%%%%%%%%%%%%%%%%%%%%%%%%%%%%%%%%%%
\subsubsection{Calcium vacancies}
%%%%%%%%%%%%%%%%%%%%%%%%%%%%%%%%%%%%%%%%%%%%%%%%%%%%%%
\label{sec:VCa}

In bulk, the Ca atom is surround by eight O atoms : four O$_1$ and four O$_2$. The distance between Ca and O$_1$ is slightly shorter than that between Ca and O$_2$. The removal of a neutral Ca atom leads to a deficit of two electrons. The four O$_1$ atoms are displaced outward by 0.02 {\AA}; the remaining four O$_2$ atoms are displaced by 0.19 {\AA}.

%The neutral charge state has $S$=1.
%Two unoccupied in-gap Kohn-Sham states are found at 0.63 eV above the VBM, and those states are almost degenerated: the energy difference is less than 0.01 eV. The holes are distributed over four of the neighboring O$_1$ atoms (Fig.~\ref{fig:fig_VCa}). 
%The four O$_1$ atoms on which holes reside are displaced outward by 0.02 {\AA}; the remaining four O$_2$ atoms are displaced by 0.19 {\AA}.

Adding electrons leads to the -- and  2-- charge states. 
In the 2-- charge state, there are no Kohn-Sham states in the gap. 
Four O$_1$ atoms move outward by 0.13 {\AA} (referenced to the interatomic distance of Ca--O in bulk), and the other four O$_2$ atoms move outward by 0.18 {\AA}.
 
\begin{figure}[]
\includegraphics[width = 6 cm]{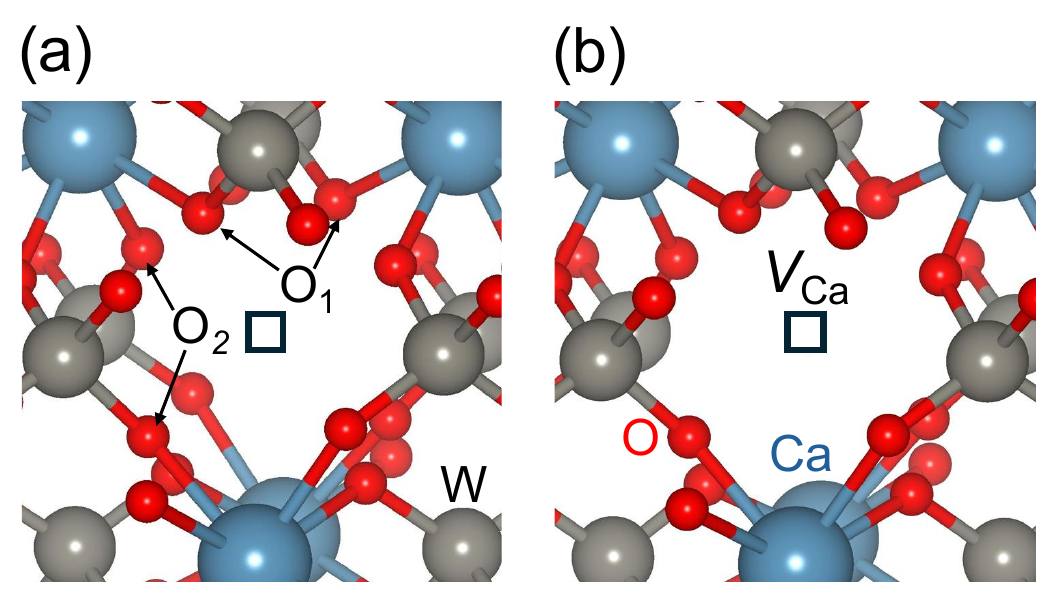}\\
\caption{\label{fig:fig_VCa} Local atomic structure for Ca vacancies in CaWO$_4$ in (a) the neutral ($V_{\rm Ca}^{0}$) and (b) the 2-- charge state ($V_{\rm Ca}^{2-}$).  %The corresponding Kohn-Sham states in the band gap, along with their occupation, are shown in the lower panels. The spin-density isosurface, representing the distribution of the unpaired holes for $V_{\rm Ca}^{0}$, is also shown for one of two degenerate in-gap states. The isosurfaces correspond to 10\% of the maximum~\cite{VESTA_11}. 
}
\end{figure}

As Fig.~\ref{fig:fig_formE} shows, only one thermodynamic transition level occurs in the band gap: a (0/2--) level at 0.46 eV~above the VBM.
%, which indicates that $V_{\rm Ca}$ is a shallow acceptor, like as in CaO~\cite{2006_zunger_cao_PRL}. 
A recent GGA+$U$ study~\cite{2023_CWO_PBEU} reported two levels, (0/--) and (--/2--),  within 0.5 eV above the VBM. 
%Although unlikely to be stable in thermodynamic equilibrium, we find that $V_{\rm Ca}^0$ possesses a triplet ground-state spin. It would therefore be good to avoid processing that puts $V_{\rm Ca}$ into a neutral charge state, as it would then contribute to magnetic noise~\cite{chatterjee_semiconductor_2021,onizhuk_decoherence_2024}.
%The triplet ground state of $V_{\rm Ca}^0$ could contribute to magnetic noise~\cite{chatterjee_semiconductor_2021,onizhuk_decoherence_2024}; however, since it is only stable when the Fermi level is below 0.46~eV, it is unlikely to occur. 

The cNEB method produced a migration barrier for $V_{\rm Ca}^{2-}$ of 1.89 eV. This corresponds to a temperature of 680 K, i.e., this defect
will be immobile at room temperature, but can diffuse during annealing.
We will comment on diffusion in Sec.~\ref{ssec:Ercomplexes}.

%We obtained $E_{\rm b}$ of 1.89 eV for , indicating that this defect might be immobile at room temperature.

%%%%%%%%%%%%%%%%%%%%%%%%%%%%%%%%%%%%%%%%%%%%%%%%%%%%%%
\subsubsection{$V_{\rm Ca} - V_{\rm O}$ complexes}
%%%%%%%%%%%%%%%%%%%%%%%%%%%%%%%%%%%%%%%%%%%%%%%%%%%%%%

%$V_{\rm Ca}$ is an important defect as much as $V_{\rm O}$, thus $V_{\rm Ca}$ and $V_{\rm O}$ could see each other in the CaWO$_4$ samples. 
Since both $V_{\rm Ca}$ and $V_{\rm O}$ are low-energy defects, particularly under W-rich conditions (Points A and B in Fig.~\ref{fig:fig_formE}), it is conceivable that they would form a complex.  The   $V_{\rm Ca} - V_{\rm O}$ complex introduces two transition levels slightly above the VBM : (2+/+)  at 0.19 eV and a (+/0)  at 0.47 eV and two levels in the upper part of the band gap: (0/--) level at 2.32 eV below the CBM and a (--/2--) level at 1.05 eV below the CBM. Thus, the defect is stable mostly in the neutral and the -- charge state.

%We examined whether formation of a $V_{\rm Ca} - V_{\rm O}$ complex would be possible.
The binding energy of the  $V_{\rm Ca} - V_{\rm O}$ complexes can be calculated as $E_{\rm bind} [(V_{\rm Ca} - V_{\rm O})^0] =   E^f (V_{\rm Ca}^{2-}) + E^f (V_{\rm O}^{2+}) -E^f [(V_{\rm Ca} - V_{\rm O})^0]$ = 1.25 eV and $E_{\rm bind} [(V_{\rm Ca} - V_{\rm O})^-] =   E^f (V_{\rm Ca}^{2-}) + E^f (V_{\rm O}^{+}) -E^f [(V_{\rm Ca} - V_{\rm O})^-]$ = 1.38 eV. 
The positive binding energy indicates that complex formation is favorable. 
The activation energy for dissociation can be estimated by adding the migration barrier for the mobile defect species to the binding energy of the complex. Using the lowest migration barrier among  $V_{\rm Ca}^{2-}$ and $V_{\rm O}^{2+}$, we would estimate an activation energy of 0.43 eV + 1.25 eV (1.38 eV)= 1.68 eV for the neutral complex and 1.84~eV for the negatively charged complex, which roughly correspond to annealing temperatures of 605 K and 662 K. 
%So, complexes can be stable at room temperature.
%and might be stable at room temperature. However, the binding is not very strong, so the complex would be dissociated at high temperature.

%%%%%%%%%%%%%%%%%%%%%%%%%%%%%%%%%%%%%%%%%%%%%%%%%%%%%%
\subsubsection{Tungsten vacancies }
%%%%%%%%%%%%%%%%%%%%%%%%%%%%%%%%%%%%%%%%%%%%%%%%%%%%%%
\label{sec:VW}

%In the case of $V_{\rm W}$, the W--O bonds are broken and the six holes (or unpaired electrons) that are left occupy O dangling bonds.  
%Removal of a W atom breaks four W--O bonds, which originally contained a total of 8 electrons; in the neutral charge state of the vacancy, 6 electrons are moved, and remaining two electrons occupy the resulting dangling bonds on the surrounding O atoms.
%More negative charge states (up to 6--) will fill these dangling bonds.

As seen in Fig.~\ref{fig:fig_formE}, we find that $V_{\rm W}$ is a deep acceptor, with a (0/2--) level at 3.32 eV above the VBM, a (2--/4--) level at 3.58 eV, and a (4--/6--) level at 3.64 eV. 
 $V_{\rm W}$  thus behaves as a ``negative-$U$'' center, in which the --, 3--, and 5-- charge states are not thermodynamically stable.
The corresponding transition levels are (0/--) = 5.23 eV, (--/2--) = 1.42 eV, (2--/3--) = 3.94 eV, (3--/4--) = 3.21 eV, (4--/5--) = 3.89 eV, and (5--/6--) = 3.39 eV (reference to the VBM).
%Therefore, the vacancy might be electrically inactive.
 Under W-rich conditions (i.e., point B and C), $V_{\rm W}$ has a very high formation energy, indicating the vacancy is unlikely to form.
 %: stable for the Fermi level position close to the CBM which is not practical. Even for the O-rich condition (i.e., point A and D), the vacancy formation may not be likely due to the relatively high formation energy near the midgap. 
At points A and D in Fig.~\ref{fig:fig_formE}, $V_{\rm W}$ the formation energy is lower, but this would require extreme O-rich conditions, which are unlikely to occur (see Sce.~\ref{sec:prev}); we thus conclude that $V_{\rm W}$ is unlikely to be present.
%These results are different from a recent study, suggesting that $V_{\rm W}$ is considerable at one growth condition ~\cite{2023_CWO_PBEU}. 
%This might be attributed to a reference phase for W chemical potentials (the reference phase is unclear) in the study, which may be thermodynamically less stable. 
We also note that the charge-state corrections~\cite{Freysoldt_09,Freysoldt_11} for the 4-- and 6-- charge states are very large, rendering the formation energies less reliable.
%; still, we think that tungsten vacancies are overall unlikely to form.

\begin{figure}[]
\includegraphics[width = 8.7 cm]{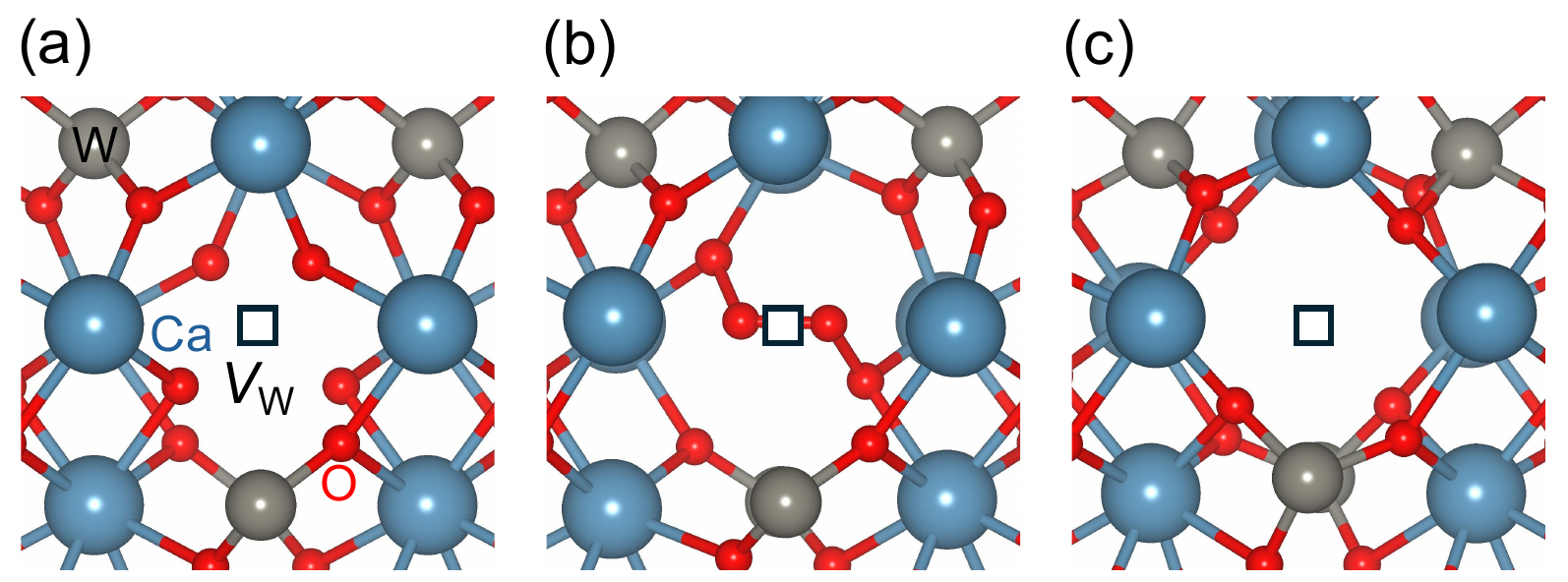}\\
\caption{\label{fig:fig_VW} Local atomic structure for W vacancies in CaWO$_4$ for (a) the unrelaxed and (b) the relaxed structure of neutral ($V_{\rm W}^{0}$) and (c) the 6-- charge state ($V_{\rm W}^{6-}$). 
%The atomic structure around $V_{\rm W}^{0}$ is significantly distorted with the surrounding O atoms moving outward when atomic relaxations allowed.
}
\end{figure}

Atomic distortion and relaxations near the vacancy are significant (Fig.~\ref{fig:fig_VW}). In the neutral charge state, two of the nearest neighboring four O atoms break two Ca-O bonds and form 
%a dumbell-like 
an O--O bond with a bond length of 1.39 \AA~at the vacancy site. Adding electrons into the supercell removes this O--O bond, and in the 6-- charge state, all four O neighbors relax outward by a large distance, 0.61~{\AA}.
 
 %In the neutral charge state, when the atomic relaxations around the vacancy are not allowed, the holes are localized on separate O atoms around $V_{\rm W}$ by introducing three empty states in the gap: one empty state at 1.19 eV and  two empty states at 1.08 eV  above the VBM.  The unrelaxed configuration (Fig.~\ref{fig:fig_VW}(a)) is higher by 5.78 eV in energy than the configuration in which the neighboring atoms around $V_{\rm W}$ are allowed to relax (Fig.~\ref{fig:fig_VW}(b)). 
%Indeed, the O atoms are significantly outward relaxed by 0.51 -- 0.71 {\AA} ~ and the local atomic structure is largely distorted. 

%Adding electrons into the supercell reduces the atomic relaxation of the neighboring atoms around  $V_{\rm W}$. The interatomic distance between  $V_{\rm W}$ and O atoms are  x.xx -- x.xx \AA in the --2 charge state and x.xx -- x.xx \AA in the --4 charge state. The configuration in the --6 charge state is bit different: three of the neighboring O atoms are relaxed by 0.31 -- 0.41 \AA and one O atom are remarkably relaxed by 1.49 \AA.

%%%%%%%%%%%%%%%%%%%%%%%%%%%%%%%%%%%%%%%%%%%%%%%%%%%%%%
\subsubsection{Oxygen interstitials}
%%%%%%%%%%%%%%%%%%%%%%%%%%%%%%%%%%%%%%%%%%%%%%%%%%%%%%
\label{sec:Oi}

%O$_i$ is one of most important defects due to its low formation energies. 
The oxygen interstitial, O$_i$, has modest formation energies, particularly under O-rich conditions (Fig.~\ref{fig:fig_formE}); as noted in Sec.~\ref{sec:prev}, highly O-rich conditions are unlikely to occur during growth.
The defect has a (0/--) level at 2.69 eV and a (--/2--) level at 3.11 eV above the VBM. ${\rm O}_i^-$ has a spin 1/2, i.e., an unpaired spin that could contribute to magnetic noise~\cite{chatterjee_semiconductor_2021,onizhuk_decoherence_2024}, while ${\rm O}_i^0$ and ${\rm O}_i^{2-}$ has spin 0 state.

%, which differs from GGA+$U$ study reporting a (0/--) level  close to the VBM~\cite{2023_CWO_PBEU}. Possibly, although no details configurations are described in Ref.~\onlinecite{2023_CWO_PBEU}, we speculate the authors consider one configuration of O$_i$ which often happened~\cite{2013_MC_LAO}.

\begin{figure}
\includegraphics[width = 6 cm]{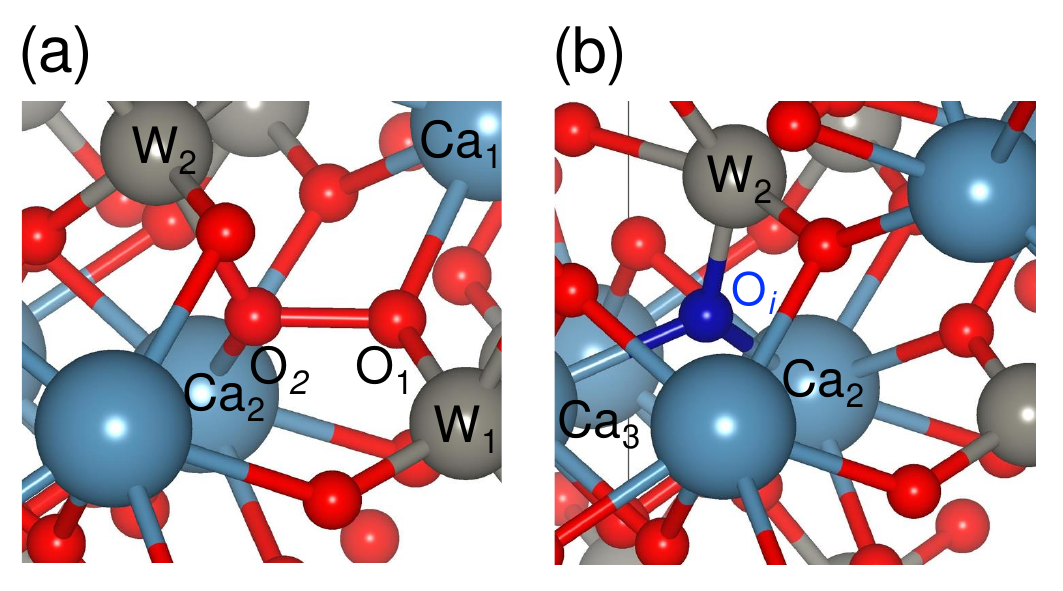}\\
\caption{\label{fig:fig_Oi}  Local atomic structure for O interstitials in CaWO$_4$ in (a) the  neutral (O$_i^{0}$)
and (b) the 2-- (O$_i^{2-}$) charge state.  O$_i^{0}$ is most stable in the split-interstitial configuration, while O$_i^{2-}$ prefers a planar configuration, bonded to one W and two Ca atoms.
}
\end{figure}

We find that O$_i$ can have two distinct configurations, as shown in Fig.~\ref{fig:fig_Oi}. 
In the neutral charge state, O$_i$ forms as a split-interstitial with an O$_1$--O$_2$ bond length of 1.43 {\AA}; similar bond lengths were reported for split-interstitial configurations in other oxides such as Al$_2$O$_3$ (1.43 \AA)~\cite{2013_MC_Al2O3_JAP} and LaAlO$_3$ (1.40 \AA)~\cite{2013_MC_LAO}. 
Labeling the atoms as in Fig.~\ref{fig:fig_Oi}, we find that O$_1$ bonds to Ca$_1$ with a bond length of 2.44 {\AA} and it also bonds to W$_1$ with a bond length of 1.89 {\AA}. The O$_2$--W$_2$ bond length is 2.17 {\AA}, and the O$_2$--Ca$_2$ bond length is 2.23 {\AA}.

In the 2-- charge state, %the O$_i$ and O$_1$ becomes separated with a distance of 2.82 {\AA}. 
O$_i$ adopts a planar configuration, having one W and two Ca nearest neighbors with a W$_2$--O$_i$ distance of 1.82 {\AA}, Ca$_2$--O$_i$ distance of 2.27 {\AA} and Ca$_3$--O$_i$ distance of 2.38 {\AA}. The -- charge state shows a similar configuration, with W$_2$-O$_i$, Ca$_2$-O$_i$, and Ca$_3$-O$_i$ distances of 1.92 {\AA}, 2.29 {\AA}, and 2.45 {\AA}, respectively.

Our study of migration led to a barrier height $E_{\rm b}$ = 0.80~eV for ${\rm O}_i^{2-}$, which corresponds to an annealing temperature of 288 K, 
%This is slightly larger than that of $V_{\rm O}^{2+}$, but still
indicating a high mobility.
If ${\rm O}_i^{2-}$ would form during growth or processing, it would most likely diffuse out of the sample, or bond to other defects or impurities.
%So, once ${\rm O}_i^{2-}$ forms in CaWO$_4$ during growth, this defect can move through the specimens and interact with the other defects and dopants.
%The O$_1$--Ca$_1$ bond is broken and the two atoms move away from each other with an  interatomic distance of 4.43 {\AA}.

%This configuration is less stable than the dumbbell configuration by 0.42 eV in the neutral charge state. %by the Coulomb interaction between O$_i^{-2}$ and the W ions in a +6 oxidation state.  

%%%%%%%%%%%%%%%%%%%%%%%%%%%%%%%%%%%%%%%%%%%%%%%%%%%%%%
\subsubsection{Calcium interstitials}
%%%%%%%%%%%%%%%%%%%%%%%%%%%%%%%%%%%%%%%%%%%%%%%%%%%%%%
\label{sec:Cai}

According to Fig.~\ref{fig:fig_formE}, Ca$_i$ has modest formation energies, particularly under Ca-rich conditions (points C and D, Fig.~\ref{chempot}). 
The defect has donor character, with a (2+/+) level at 1.48 eV below the CBM and a (+/0) level at 0.31 eV below the CBM.
%, which are similar to a previous report~\cite{2023_CWO_PBEU}. 
%Therefore, the defect easily releases one electron into the conduction band. I.e., Ca$_i$ is a shallow donor.

\begin{figure}
\includegraphics[width = 6 cm]{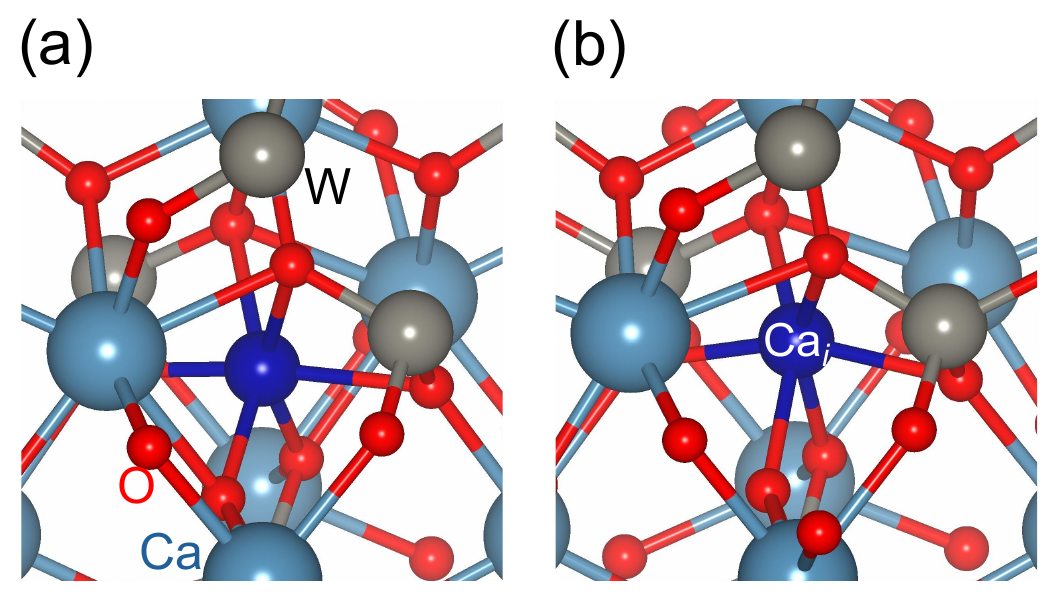}\\
\caption{\label{fig:fig_Cai}  Local atomic structure for Ca interstitials in CaWO$_4$ in (a) the neutral (Ca$_i^{0}$)
and (b) the 2+ (Ca$_i^{2+}$) charge state.  
}
\end{figure}

As shown in Fig.~\ref{fig:fig_Cai}, Ca$_i$ sits on an interstitial site with six neighboring O atoms.
Three of the Ca$_i$-O bond lengths are distinctly shorter than the Ca--O bond length in pristine CaWO$_4$ (Table~\ref{param}): 
%the computed $d_{\rm Ca_i-O}$ are 
2.16 {\AA}, 2.30 {\AA},~and 2.31 {\AA} in the neutral charge state, and 2.22 {\AA},  2.28 {\AA},~and 2.18 {\AA} in the 2+ charge state. 

%Chris 051326
We calculated the migration barrier of ${\rm Ca}_i^{2+}$, finding a low barrier of $E_{\rm b}$ = 0.51 eV.
The migration proceeds via an interstitialcy mechanism, in which the interstitial atom moves toward a substitutional Ca atom, pushes this atom out of the lattice site, and takes its position; the atom that was originally on the lattice site is forced into a new interstitial position.
The barrier of $E_{\rm b}$ = 0.51 eV corresponds to an annealing temperature of about 187~K, and ${\rm Ca}_i^{2+}$ is thus highly mobile.

%We calculated the migration barrier of ${\rm Ca}_i^{2+}$, finding $E_{\rm b}$ = 2.10 eV, which corresponds to an annealing temperature of about 760~K.

%%%%%%%%%%%%%%%%%%%%%%%%%%%%%%%%%%%%%%%%%%%%%%%%%%%%%%
\subsubsection{Tungsten interstitials}
%%%%%%%%%%%%%%%%%%%%%%%%%%%%%%%%%%%%%%%%%%%%%%%%%%%%%%
\label{sec:Wi}

\begin{figure}
\includegraphics[width = 6 cm]{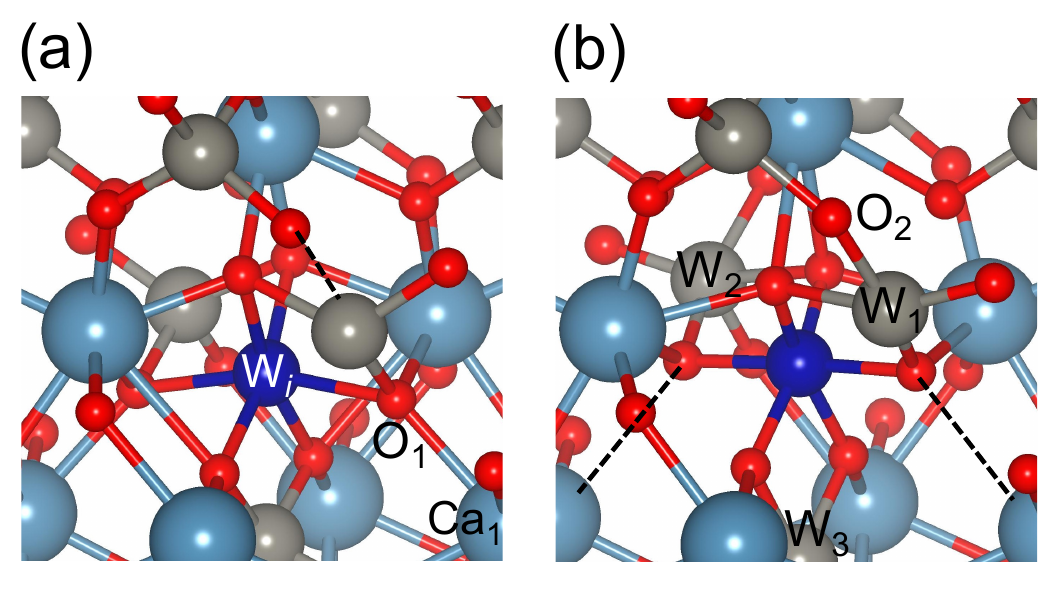}\\
\caption{\label{fig:fig_Wi} Local atomic structure for W interstitials in CaWO$_4$ in (a) the neutral (W$_i^{0}$)
and (b) the 6+  charge state (W$_i^{6+}$).  
}
\end{figure}

The tungsten interstitial, W$_i$, has five charge-state transition levels in the gap: (6+/4+) at 1.35 eV, (4+/3+) at 2.36 eV, (3+/2+) at 3.03 eV, (2+/+) at 3.43 eV, and (+/0) at 4.09 eV above the VBM.  However, W$_i$ is unlikely to form, since its formation energy is high even under W-rich conditions (Fig.~\ref{fig:fig_formE}).

In the neutral charge state, W$_i$ is positioned at a site surrounded by six O atoms with three distinct bond lengths: $d_{\rm {W}_i-\rm {O}}$ = 2.07 \AA, 2.13 \AA,~ and 2.16 \AA ~[Fig.~\ref{fig:fig_Wi}(a)]. 
W$_i$ is quite close to three W atoms, labeled W$_1$, W$_2$, and W$_3$ in the figure: $d_{\rm {W}_i-\rm {W}_1}$ = 2.66 \AA, $d_{\rm {W}_i-\rm {W}_2}$ = 2.66 \AA, and $d_{\rm {W}_i-\rm {W}_3}$ = 2.49 \AA ; these distances are much shorter than the W--W distance (3.88 \AA) in pristine CaWO$_4$ (Table~\ref{param}). 
This leads to W $5d-5d$ coupling and introduces three occupied Kohn-Sham states above the VBM, at 2.05 eV,  2.45 eV, and 3.22 eV. 

The local atomic structure is sensitive to the charge state of W$_i$.
%, and change in the charge state cause strong distortions.  
In the 6+ charge state [Fig.~\ref{fig:fig_Wi}(b)], the $d_{\rm {W}_i-\rm {O}}$ distances are reduced to 1.86 {\AA} ($\times$2), 1.95 {\AA} ($\times$2), and 1.92 {\AA} ($\times$2), consistent with the fact that as the ion charge increases, the ion radius decreases~\cite{Shannon}. 
The W$_i$--W distance becomes 3.02 {\AA} for W$_1$, 3.02 {\AA} for W$_2$, and 2.94 {\AA}  for W$_3$ [Fig.~\ref{fig:fig_Wi}(b)].
%Ca$_1$ and O$_1$ ($d_{\rm {Ca}_1-\rm {O}_1}$ = 2.48 \AA) atoms are separated by an interatomic distance of 3.07 \AA, whereas W$_1$ and O$_2$ atoms get closer: $d_{\rm {W}_1-\rm {O}_2}$ = 2.68 {\AA} for W$_i^0$ and 1.97 {\AA} for W$_i^{6+}$. 

%%%%%%%%%%%%%%%%%%%%%%%%%%%%%%%%%%%%%%%%%%%%%%%%%%%%%%
\subsubsection{Antisites}
%%%%%%%%%%%%%%%%%%%%%%%%%%%%%%%%%%%%%%%%%%%%%%%%%%%%%%

The Ca$_{\rm W}$ antisite has three acceptor levels in the gap: (0/--) at 2.27 eV, (--/3--) at 2.53 eV, (3--/4--) at 2.85 eV above the VBM. In the neutral charge state, although each W atom in pristine CaWO$_4$  is bonded to four neighboring O atoms with the same bond length (Table~\ref{param}), the Ca atom substituting on the W site distorts the local atomic structure around the Ca atom: $d_{\rm {Ca_{W}}-\rm {O}}$ = 2.20~\AA, 2.46~\AA, 2.49~\AA, and 2.21~\AA. This may be attributed to the much larger ionic radius of fourfold-coordinated Ca(II) (1.00~{\AA} {\it vs.} 0.42~{\AA} for W(VI)~\cite{Shannon}). Adding electrons renders the Ca$_{\rm W}$--O bond lengths more uniform, and in the 4-- charge state the four Ca$_{\rm W}$--O bond have the same bond length of 2.26~{\AA}.

The W$_{\rm Ca}$ antisite, a W atom substituting on the Ca site, introduces four defect levels in the gap: (4+/3+) = 0.86 eV, (3+/2+) = 2.52 eV, (2+/+) = 3.50 eV, and (+/0) = 4.95 eV above the VBM. 
Although each Ca atom in CaWO$_4$ is surrounded by eight O atoms, W$_{\rm Ca}$ is bonded to four neighboring O atoms with $d_{\rm {W_{Ca}}-\rm {O_1}}$ = 2.17 \AA~ in the neutral charge state. This bond length is much larger than the W--O bond length in pristine CaWO$_4$ (Table~\ref{param}). The $\rm {W_{Ca}}-\rm {O_2}$ distance with the next four oxygen neighbors is 2.40 \AA. 
As the magnitude of the defect charge increases, $d_{\rm {W_{Ca}}-\rm {O_1}}$ decreases: 2.16 \AA~ in the + charge state, 2.09 \AA~ in the + charge state, and 2.10 \AA~in the 3+ charge state. In the 4+ charge state $\rm {W_{Ca}}$ is found to bond to eight O atoms, and the second set of O atoms now get closer than the first set of O atoms in other charge states: $\rm {W_{Ca}}$--$\rm {O_1}$ = 2.18 \AA~and  $\rm {W_{Ca}}$--$\rm {O_2}$ = 1.96 \AA.

Both antisites  (W$_{\rm Ca}$ and Ca$_{\rm W}$) are however unlikely to form in CaWO$_4$ (Fig.~\ref{fig:fig_formE}) since their formation energies are very high (except under extreme O-rich conditions, which are unlikely to occur; see Sec.~\ref{sec:prev}).

%%%%%%%%%%%%%%%%%%%%%%%%%%%%%%%%%%%%%%%%%%%%%%%%%%%%%%
\subsection{Prevalent defects}
%%%%%%%%%%%%%%%%%%%%%%%%%%%%%%%%%%%%%%%%%%%%%%%%%%%%%%
\label{sec:prev}

\begin{figure}[]
\includegraphics[width = 6 cm]{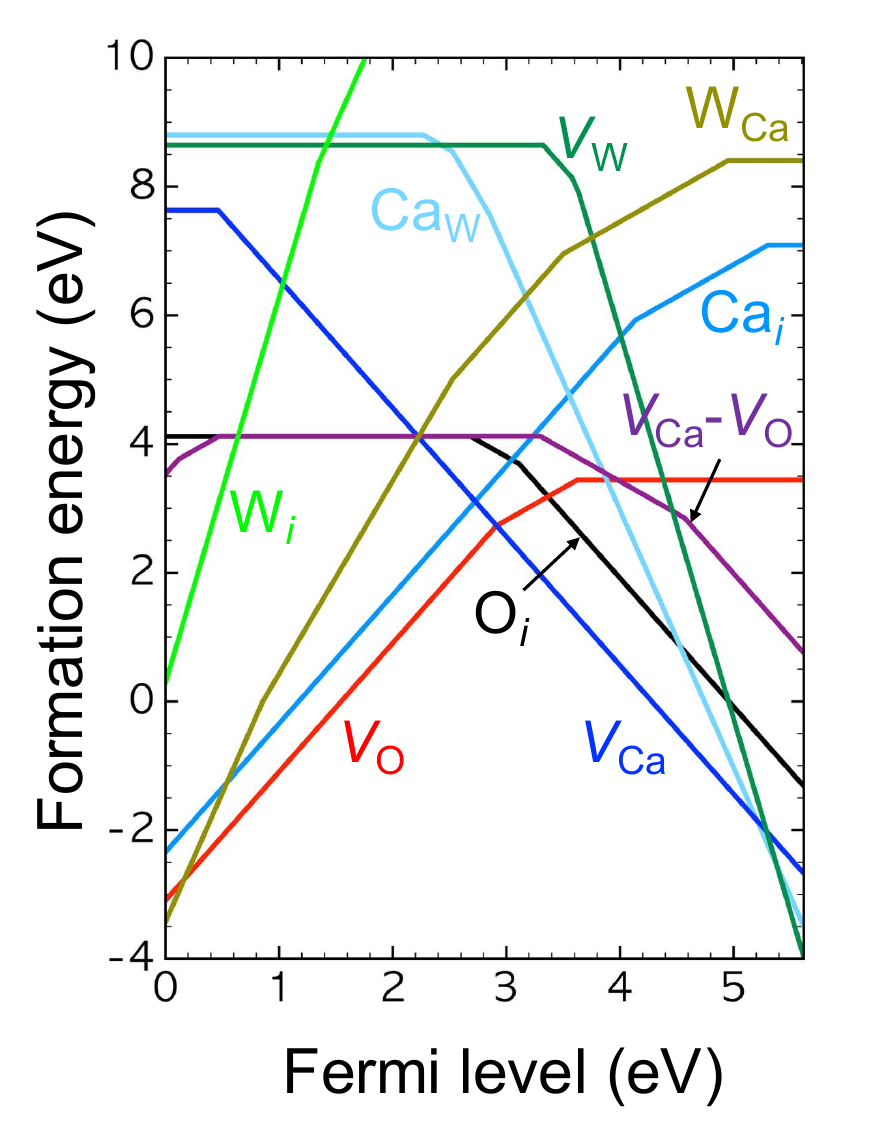}\\
\caption{ \label{fig:fig_formE_MBE}  Formation energies of native point defects as a function of the Fermi level at conditions reflecting MBE or PLD growth (see text).}
\end{figure}

In the preceding sections we discussed all possible native defects and their formation energies. We emphasized these formation energies (and hence the likelihood of incorporating the defect during growth or annealing) depend on Ca, W, and O chemical potentials, and in Fig.~\ref{fig:fig_formE} we presented four sets of formation energies for limiting chemical-potential conditions (as identified in Fig.~\ref{chempot}).
We now discuss, based on available information about experimental growth conditions, which conditions are most likely to occur.
This is difficult to assess quantitatively for bulk single-crystal growth, but
CaWO$_4$ films have also been grown using techniques such as pulsed laser deposition (PLD) and molecular beam epitaxy (MBE). 
%It is difficult to know the values of chemical potentials during growth.  
As an estimate for the conditions during PLD or MBE growth we set the oxygen chemical potential to $\mu_{\rm O}$ = --1.84~eV, corresponding to
an O$_2$ partial pressure $P_{\rm O_2}$ = 10$^{-6}$ mbar at a temperature $T$ = 1000 K~\cite{2013_MC_LAO}. 
These values lie in the range of typical growth conditions for PLD or MBE ($P_{\rm O_2}$ =10$^{-3}$--10$^{-7}$ mbar and $T$ = 900--1100 K).~\cite{Kaur_PLD_2021,Tang_MBE_2024}  
We also set $\mu_{\rm W}$=--3.41 eV, which corresponds to the midpoint between Ca-rich and W-rich conditions at $\mu_{\rm O}$ = --1.84~eV.
These chemical-potential conditions are indicated by the symbol ``x'' in Fig.~\ref{chempot} and the corresponding formation energies are plotted in Fig.~\ref{fig:fig_formE_MBE}.

%We also set $\mu_{\rm W}$=--2.66 eV, which corresponds to taking WO$_3$ as the limiting phase. 

%CaWO$_4$ single crystals are mostly grown by the Czochralski method~\cite{Erb_C2CE_2013,SILVA_2014}, whose main disadvantage is contamination with the crucible material.  Due to the high melting point ($\sim$1600 $^{\circ}$C), a low oxygen partial pressure is used to minimize the reaction between the crucible material and oxygen~\cite{Erb_C2CE_2013}. Therefore the oxygen chemical potential for single-crystal growth is likely to be lower than the $\mu_{\rm O}$ = --1.84~eV value assumed above.
% may far away from the oxygen-rich condition (i.e., $\mu_{\rm O}$ = 0 eV).  

\begin{figure*}
\includegraphics[width = 14 cm]{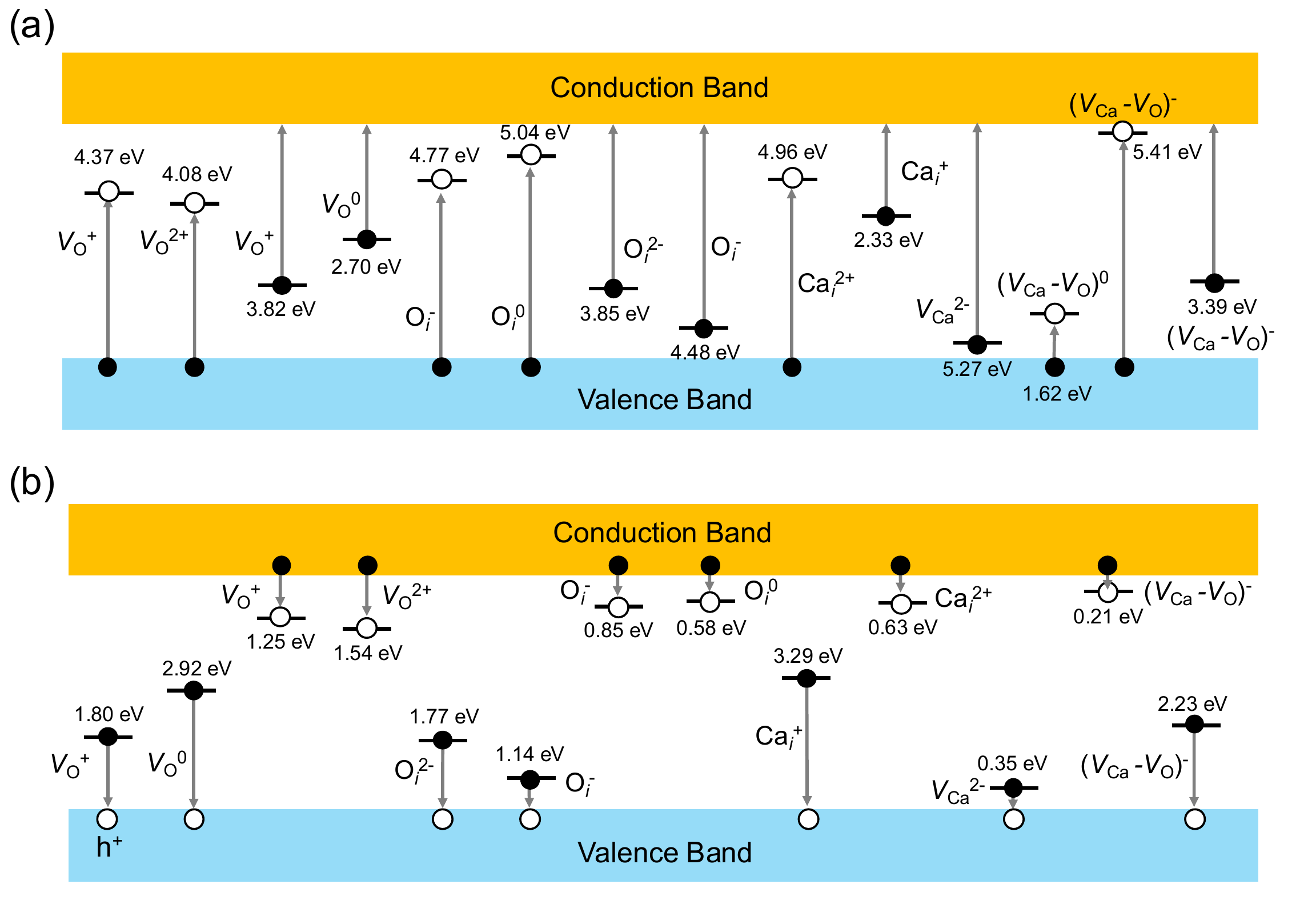}\\
\caption{\label{fig:fig_opt}Microscopic mechanisms for (a) absorption and (b) luminescence processes associated with native defects in CaWO$_4$.  Note that the depicted levels are not thermodynamic transition levels (zero-phonon lines) but correspond to peaks in absorption or emission spectra.
}
\end{figure*}

For the conditions assumed in Fig.~\ref{fig:fig_formE_MBE}, we can also estimate where the Fermi level is expected to be situated based on the requirement of charge neutrality between acceptor- and donor-like defects.  Since defect concentrations depend exponentially on formation energies, the lowest-energy defects are dominant, and positive and negative charges need to balance.
Figure~\ref{fig:fig_formE_MBE} then shows that oxygen vacancies and calcium vacancies are most likely to form, and that the Fermi level will be positioned in the vicinity of 2.9~eV, i.e., near mid-gap, which is consistent with the CaWO$_4$ samples being insulating.

A notable finding is that W-related defects, i.e., W vacancies ($V_{\rm W}$), W interstitials (W$_i$), and antisites (W$_{\rm Ca}$ and Ca$_{\rm W}$) are unlikely to form since their formation energies are high for the expected range of Fermi levels.

Going back to Fig.~\ref{fig:fig_formE}, we can examine what conditions would be required to cause a significant deviation from this conclusion.
Defects other than $V_{\rm O}$ and $V_{\rm Ca}$ would be present under more extreme O-rich conditions, which (as argued above) are unlikely to occur.
Even with $V_{\rm O}$ and $V_{\rm Ca}$ dominating, the Fermi level could shift due to differing growth conditions.
More O-rich conditions will promote formation of $V_{\rm Ca}$ and shift the Fermi to lower values, while more O-poor conditions will shift the Fermi to higher values.
Oxygen-poor conditions, with chemical potentials lower than the $\mu_{\rm O}$ = --1.84~eV value assumed above, are more likely to occur during single-crystal growth.
CaWO$_4$ single crystals are mostly grown by the Czochralski method~\cite{Erb_C2CE_2013,SILVA_2014}, in which a low oxygen partial pressure is used to minimize the reaction between the crucible material and oxygen~\cite{Erb_C2CE_2013} in order to avoid contamination with the crucible material~\cite{Erb_C2CE_2013}. 
Still, given the range of achievable growth conditions, the Fermi level is unlikely to shift far from midgap (i.e., outside the range from 2~eV to 4~eV), unless the material is deliberately doped. 

We note that these considerations about Fermi-level position were premised on the absence of any impurities.  Electrically active impurities could of course shift the Fermi level, but they are still unlikely to stimulate formation of defects other than $V_{\rm O}$ and $V_{\rm Ca}$.
We also note that under conditions that significantly deviate from equilibrium, such as ion implantation, defects other than $V_{\rm O}$ and $V_{\rm Ca}$ could of course be formed.  However, since implantation is usually followed by annealing (to reduce implantation damage), the assumption of equilibrium may well be approximately satisfied, particularly since our calculated values for migration barriers for relevant defects are all modest and hence significant diffusion is possible.

%well away from the band edges (near $\sim$3 eV, where the formation energy of Al vacancies becomes lower than that of O vacancies) owing to consistent with the fact that CaWO$_4$ is an insulator. The stable defects are Ca vacancies, Ca interstitials, O vacancies, and O interstitials. As explained above, the W-related defects are unlikely to be present.

%The stable charge states of Ca vacancies and  Ca interstitials are 2-- and 2+, respectively, over a wide ranges of Fermi levels near midgap, On the other hand, those of O vacancies and O interstitials may more depends on growth conditions because all their defect levels are positioned near the midgap. For the position of the Fermi level near $\sim$3 eV, O vacancies are stable in the positive and  neutral charge states and O interstitials is in the 2-- charge state.

%Overall, O-related defects ($V_{\rm O}$ and O$_i$) and Ca-related defects ($V_{\rm Ca}$ and Ca$_i$) are the most likely to form.
%For most of the conditions, two acceptor-like defects, $V_{\rm Ca}$ and O$_i$ are the lower-energy defects in the higher part of the band gap, while  donor-like defects ($V_{\rm O}$ and Ca$_i$) become important in the lower part of the band gap. 

%At point B, two interstitials, Ca$_i$ and  O$_i$ are the predominant defects. At point A', two donors, $V_{\rm O}$ and Ca$_i$ have low formation energies. At point B', Ca$_i$ and  O$_i$ are most likely to form and the crossing point of the formation energies for the two defects is 3.13 eV, where is nearly at the mid-gap. 

%%%%%%%%%%%%%%%%%%%%%%%%%%%%%%%%%%%%%%%%%%%%%%%%%%%%%%
\subsection{Optical transitions}
%%%%%%%%%%%%%%%%%%%%%%%%%%%%%%%%%%%%%%%%%%%%%%%%%%%%%%
\label{sec:opt}

For the defects that are expected to be most likely to occur, we also investigated optical transition energies for excitations involving carriers in either the VBM or the CBM.
In addition to $V_{\rm O}$ and $V_{\rm Ca}$, O$_i$ and Ca$_i$ also have relatively low formation energies.
%$V_{\rm Ca}$ only has transition levels close to the VBM, and hence is unlikely to produce optical transitions in the visible.
We determine possible optical transitions associated with these defects based on the calculated formation energies and using the Frank-Condon principle~\cite{RMP_2014}.
The results are shown in Fig.~\ref{fig:fig_opt}.  

%The transition $V_{\rm O}^{+} \rightarrow V_{\rm O}^{2+} + e^-$ give rises to an absorption peak at 3.82 eV, and  $V_{\rm O}^{0} \rightarrow V_{\rm O}^{+} + e^-$ at 2.70 eV. The transition ${\rm O}_i^{2-} \rightarrow {\rm O}_i^{-} + e^-$ results in absorption at 3.85 eV. 
%These values are close to the experimentally observed at 340 nm (3.65 eV) and 520 nm (2.38 eV) absorption bands in crystals~\cite{Baccaro_pssa_2000} and 25000 cm$^{-1}$ (3.10 eV) and 32000 cm$^{-1}$ (3.97 eV) peak in crystals~\cite{YAKOVYNA_2004}. The transition $(V_{\rm Ca} - V_{\rm O})^- \rightarrow (V_{\rm Ca} - V_{\rm O})^0 + e^-$ give rises to an absorption peak at 3.39 eV, which is also close to above the experimentally reported values.

Experimentally, absorption peaks have been observed in CaWO$_4$ crystals at 340 nm (3.65 eV) and 520 nm (2.38 eV)~\cite{Baccaro_pssa_2000} and at 25000 cm$^{-1}$ (3.10 eV) and 32000 cm$^{-1}$ (3.97 eV)~\cite{YAKOVYNA_2004}. 
The absorption peaks at 3.65~eV and 3.97 eV match reasonably well with the transition $V_{\rm O}^{+} \rightarrow V_{\rm O}^{2+} + e^-$, for which we calculate an absorption peak at 3.82 eV.
As noted in Sec.~\ref{sec:VO}, the oxygen vacancy is highly mobile, so it is unlikely to occur as an isolated defect.  
$V_{\rm O}$ is more likely to bind to an acceptor,  forming a complex; the transition energy in this complex will be shifted somewhat from that in the isolated $V_{\rm O}$. 
%it might cause some optical transitions in experiment. 
For instance, for the $(V_{\rm Ca} - V_{\rm O})$ complex we calculate that the transition $(V_{\rm Ca} - V_{\rm O})^- \rightarrow (V_{\rm Ca} - V_{\rm O})^0 + e^-$ gives rises to an absorption peak at 3.39 eV, which is close to an experimentally reported peak at 3.10 eV~\cite{YAKOVYNA_2004}. Another transition associated with the complex, the transition $(V_{\rm Ca} - V_{\rm O})^0  \rightarrow (V_{\rm Ca} - V_{\rm O})^-  + h^+$ provides an absorption peak at 1.62 eV.

Alternatively, the absorption energy peaks at 3.65~eV and 3.97 eV might be associated with ${\rm O}_i$ (or a related complex); ${\rm O}_i^{2-} \rightarrow {\rm O}_i^{-} + e^-$ gives a peak at 3.85 eV. This assignment would be consistent with the experimental observation that reports a decrease in absorption at 3.10 eV and an increase at 3.97 eV when the sample is annealed in air and oxygen~\cite{YAKOVYNA_2004}. 

%We also suggest that the 2.38~eV absorption could be due to  ${\rm Ca}_i^{+} \rightarrow {\rm Ca}_i^{2+} + e^-$ at 2.23 eV. The transition $V_{\rm O}^{0} \rightarrow V_{\rm O}^{+} + e^-$ at 2.70 eV is also close to the peak at 2.38 eV, but the oxygen vacancy is highly mobile, so it might be unlikely to occur.
The 2.38~eV absorption could be due to $V_{\rm O}^{0} \rightarrow V_{\rm O}^{+} + e^-$, calculated at 2.70 eV, which could be somewhat shifted due to complex formation with an acceptor.
Alternatively, it could be assigned to ${\rm Ca}_i^{+} \rightarrow {\rm Ca}_i^{2+} + e^-$ at 2.33 eV, but under the conditions depicted in Fig.~\ref{fig:fig_formE_MBE},  ${\rm Ca}_i$ is less likely to be present than $V_{\rm O}$.

 %The transition $V_{\rm Ca}^{-}+e^-\rightarrow V_{\rm Ca}^{2-}$ gives rise to an emission peak at 5.11 eV.  
 
Optical excitation can also give rise to luminescence, for instance when a carrier returns to the defect level that was excited.
Alternatively, above-band-gap excitation can create electron-hole pairs, and the electrons or holes could be captured by defects in their equilibrium charge state.
Experimentally, photoluminescence spectra of CaWO$_4$ have produced emission peaks at 450 nm (2.76 eV) and 550 nm (2.25 eV) in powders~\cite{Campos_2007},  437 nm (2.84 eV), 570 nm (2.18 eV) in nanoparticles~\cite{KAUR_nano_2020} and 402 nm (3.08 eV), 408 nm (3.04 eV), 430 nm (2.88 eV), 498 nm (2.49 eV) in thin films~\cite{Kaur_PLD_2021}. It was also reported that a CaWO$_4$ single crystal grown by a laser-heated pedestal technique exhibited an emission peak at 420 nm (2.95 eV)~\cite{SILVA_2014}.

%These peaks in Refs.~\onlinecite{Campos_2007,Kaur_PLD_2021} were also found to be in CaMoO$_4$, possibly consistent with oxygen -related defects being the source.

The transition $(V_{\rm Ca} - V_{\rm O})^- + h^+\rightarrow (V_{\rm Ca} - V_{\rm O})^0$ provides an emission energy of 2.23 eV, which could potentially explain the emission observed at lower energies, such as 2.25 eV~\cite{Campos_2007} or 2.18 eV~\cite{KAUR_nano_2020}.
%, and 2.49 eV~\cite{Kaur_PLD_2021}.

%We expect holes to be rapidly nonradiatively captured by Ca vacancies, since their transition level is so close to the VBM, changing their charge state from 2-- to --.  
%An electron in the CBM can then radiatively recombine with the trapped hole.  

As for the emission observed at higher energies, we suggest that 
Ca interstitials, which are initially in the 2+ charge state, could be excited to ${\rm Ca}_i^{+}$ by capturing an electron.  
The subsequent transition ${\rm Ca}_i^{+}+h^-\rightarrow {\rm Ca}_i^{2+}$ results in an emission peak at 3.29 eV.
This transition might correspond to the observed higher-energy emission~\cite{Campos_2007,KAUR_nano_2020,Kaur_PLD_2021,SILVA_2014}. 
Alternatively, oxygen vacancies might be involved.
The $V_{\rm O}^0 + h^+\rightarrow V_{\rm O}^+$ transition leads to emission at 2.92~eV.  As discused above, since oxygen vacancies are quite mobile they are unlikely to occur in isolated form, but a complex with an acceptor may still emit at similar energies, potentially explaining the emission at 2.8 --3.1 eV~\cite{Campos_2007,KAUR_nano_2020,Kaur_PLD_2021,SILVA_2014}.  

\subsection{Erbium}
\label{sec:Er}

%%%%%%%%%%%%%%%%%%%%%%%%%%%%%%%%%%%%%%%%%%%%%%%%%%%%%%

\subsubsection{Substitutional erbium}

\begin{figure}[]
\includegraphics[width = 6 cm]{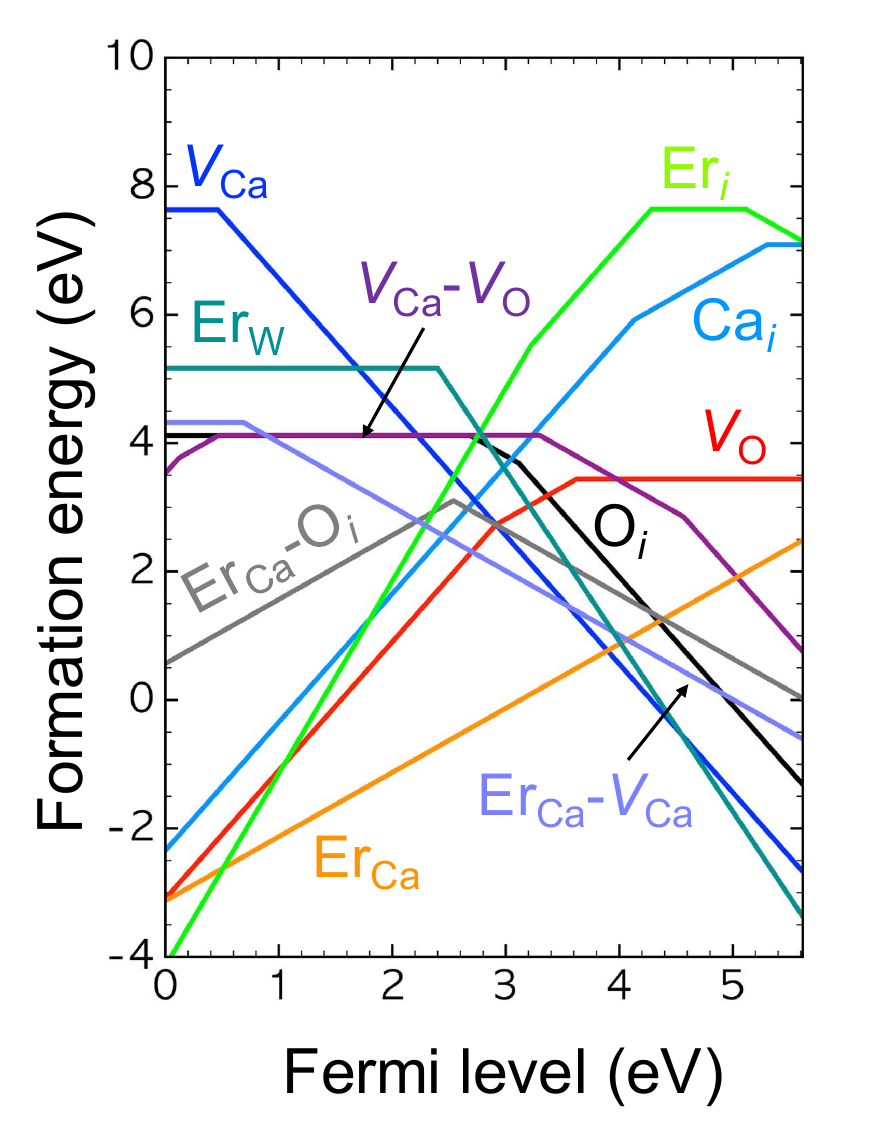}\\
\caption{ \label{fig:fig_formE_ErMBE}  Formation energies of Er dopant and its complexes with $V_{\rm Ca}$ and O$_i$ as a function of the Fermi level at conditions reflecting MBE and PLD growth. Several important native defects are also included for comparison.}
\end{figure}

CaWO$_4$ is a suitable host for rare earth ions in quantum devices, in particular because Er is incorporated on a non-polar Ca site, reducing the sensitivity to charges elsewhere in the crystal
%thus increasing the coherence time
and thus reducing the spectral diffusion of the optical transition~\cite{2023_Nature_Jeff_ErCWO}. 
%Hence, substitutional Er (Er$_{\rm Ca}$) is positively charged, i.e., Er$_{\rm Ca}^{+}$. 
%Our results confirm that Er substitutes on the Ca site with a positive charge state, which is a high-symmetry lattice site with$C_{4h}$ symmetry (see Figs.~\ref{fig:fig_formE_Er} and ~\ref{fig:fig_complex}(a)).
Our results for substitutional Er in CaWO$_4$ (Er$_{\rm Ca}$) are shown in Fig.~\ref{fig:fig_formE_ErMBE}, again for the chemical-potential conditions discussed in Sec.~\ref{sec:prev}.
The formation energies are very low, indicating that Er$_{\rm Ca}$ will easily incorporate.
Er$_{\rm Ca}$ is in the positive charge state over the entire band gap, consistent with the fact that Er prefers to be in a 3+ oxidation state and is substituting on a Ca site with valence 2.
%A (+/0) transition level occurs at 0.36 eV below the CBM.
%The fact that this (+/0) level is so far from the VBM implies that very high energy photons would be needed to change the charge state; Er$_{\rm Ca}$ will thus be immune to photoionization.

%\begin{figure*}
%\includegraphics[width = 16 cm]{fig_formE_Er}\\
%\caption{ \label{fig:fig_formE_Er}  Formation energies of Er dopant and its complexes with $V_{\rm Ca}$ and O$_i$ as a function of the Fermi level at four representative conditions as specified in Fig.~\ref{chempot} and Table~\ref{enthalpy}. The results for $V_{\rm Ca}$ and O$_i$ and included for completeness.}
%\end{figure*}

Erbium on the substitutional W site, Er$_{\rm W}$, introduces a transition level (0/3--) at 2.40 eV above the VBM.  However, Er$_{\rm W}$ has much higher formation energies, indicating that  Er$_{\rm W}$ is unlikely to form, similar to the conclusion we reached for W-related native defects.
%(--/2) at 0.25 eV, (2--/3--) at 2.34 eV

\begin{figure}[]
\includegraphics[width = 7 cm]{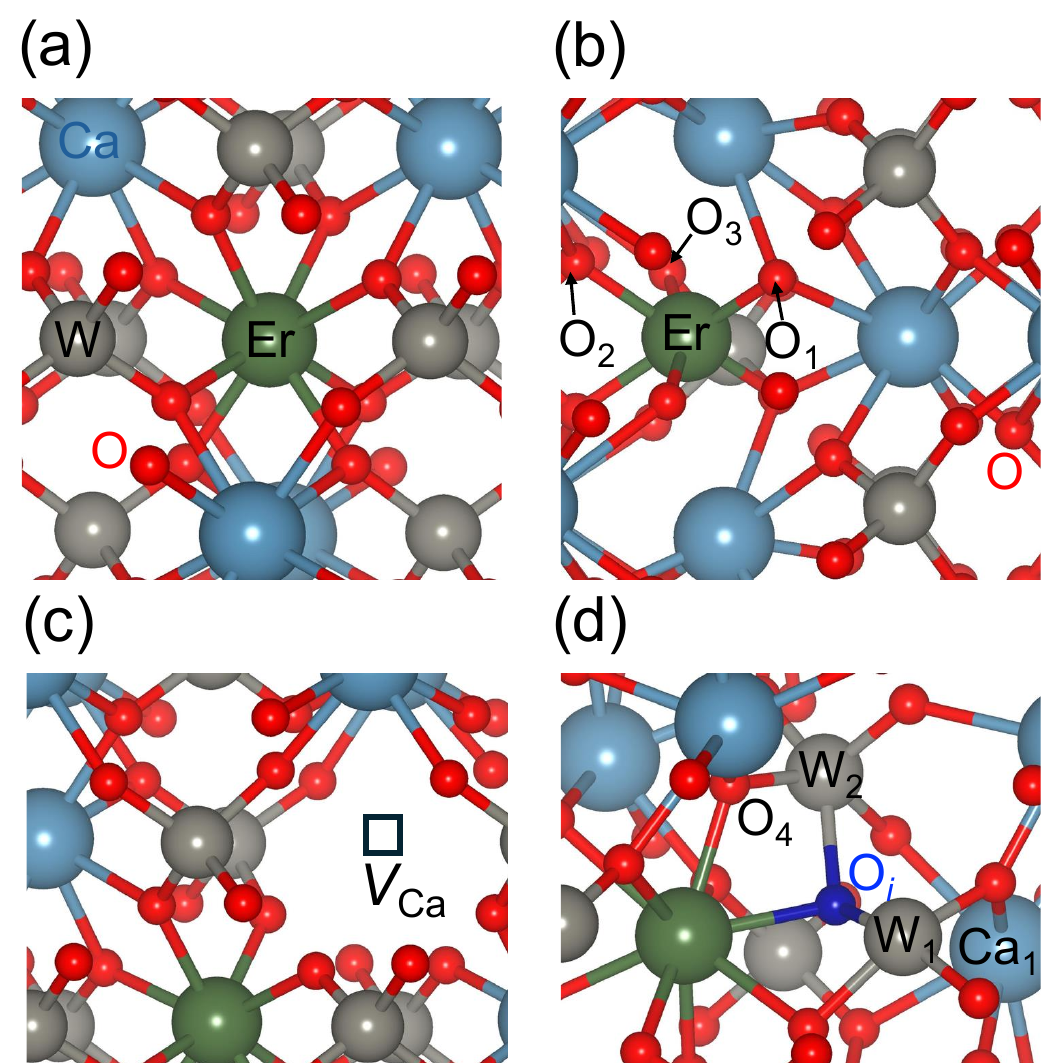}\\
\caption{\label{fig:fig_complex} Local atomic structure for (a) ${\rm Er_{Ca}}^+$, (b) ${\rm Er}_i^{3+}$, (c) $({\rm Er_{Ca}}-V_{\rm Ca})^-$,  and (d) $({\rm Er_{Ca}}-{\rm O}_i)^-$.  
}
\end{figure}

The atomic structure of Er$_{\rm Ca}^+$  is shown in Fig.~\ref{fig:fig_complex}(a). 
The neighboring O atoms relax inward, leading to Er--O bond lengths of 2.37 {\AA} and 2.38  {\AA} for Er$_{\rm Ca}^+$, shorter than the 2.48 {\AA} and 2.45 {\AA} Ca--O bond length in pristine CaWO$_4$ (Table~\ref{param}).  The corresponding Er--O bond lengths for Er$_{\rm Ca}^0$ are 2.35  {\AA} and 2.38 {\AA}.
We checked off-centered configurations where Er is moved away from the substitutional Ca site, but the Er atom prefers to move back to the Ca site.
%Er$_{\rm Ca}$ thus retains high symmetry ($C_{4h}$). 
%Our calculations thus confirm that Er occupies a high-symmetry lattice site with $S_4$ symmetry. This point group contains inversion symmetry, indicating that the site will have reduced sensitivity to charge noise as previously suggested~\cite{2023_Nature_Jeff_ErCWO}.
Our calculations thus confirm that Er occupies a high-symmetry lattice site with $S_4$ symmetry. This point group is non-centrosymmetric but does not allow a permanent electric dipole moment, indicating that the site will have reduced sensitivity to charge noise as previously suggested~\cite{2023_Nature_Jeff_ErCWO}.

%%%%%%%%%%%%%%%%%%%%%%%%%%%%%%%%%%%%%%%%%%%%%%%%%%%%%%
\subsubsection{Erbium interstitials}
%%%%%%%%%%%%%%%%%%%%%%%%%%%%%%%%%%%%%%%%%%%%%%%%%%%%%%
\label{sssec:Eri}

The formation energies of  Er interstitial (Er$_i$) are relatively high (Fig.~\ref{fig:fig_formE_ErMBE}), indicating that Er$_i$ will not easily incorporate during growth.
However, interstitials could still form when implantation is used to introduce erbium. 
The Er interstitial has three levels in the gap: (3+/2+) at 2.39 eV, (2+/0) at 1.34 eV, (0/--) at 0.50 eV below the CBM.
% and the 3+ charge state is stable in the most of the band gap. 

%The atomic structure of Er interstitials is shown in Figs.~\ref{fig:fig_Eri}(a) and (b). 

The atomic structure of Er interstitials in the 3+ charge state is shown in Fig.~\ref{fig:fig_complex}(b). The Er atom is coordinated by six neighboring O atoms, forming three distinct bond lengths. The Er--O bond lengths remain nearly unchanged upon ionization: 2.16 {\AA} (O$_1$) , 2.18 {\AA} (O$_2$), and 2.21 {\AA} (O$_3$) in the neutral state, compared to 2.16 {\AA} (O$_1$), 2.14 {\AA} (O$_2$), and 2.21 {\AA} (O$_3$)  in the 3+ charge state. In contrast, they change dramatically in the negatively charged state, particularly for the intermediate bonds, which become 2.13 {\AA} (O$_1$), 2.56 {\AA} (O$_2$), and 2.21 {\AA} (O$_3$).

%\begin{figure}[]
%\includegraphics[width = 9 cm]{fig_Eri}\\
%\caption{\label{fig:fig_Eri} Local atomic structure for (a) ${\rm Er}_i^{3+}$, (a) ${\rm Er}_i^{-}$,  and (c) $({\rm Er}_i-V_{\rm Ca})^+$.  }
%\end{figure}

%Chris 051326
Our calculations for migration of ${\rm Er}_i^{3+}$ showed that direct motion from one interstitial site to another would require overcoming a barrier of $E_{\rm b}$ = 2.32 eV.
However, the interstitialcy mechanism discussed in Sec.~\ref{sec:Cai} for Ca$_i$ also applies to Er interstitials: we find that ${\rm Er}_i^{3+}$ can exchange places with a Ca host atom with an energy barrier of only 0.94~eV, a process that could therefore already occur at temperatures slightly above room temperature.  Once the Er interstitial ``kicks into'' the lattice it will be stable there, and the Ca host atom that is ejected will form a Ca$_i^{2+}$ that will rapidly move away.  
Further diffusion of Er would require assistance by point defects.
In principle this could happen assisted by Ca interstitials, and
our calculations show that the barrier for a Ca$_i^{2+}$ to kick an Er$_{\rm Ca}$ atom out of a lattice site is 1.06 eV.
Such a low barrier would in principle favor interstitial-assisted diffusion, but the Coulomb repulsion between Ca$_i^{2+}$ and Er$_{\rm Ca}^{+}$ renders this less likely.

%%%%%%%%%%%%%%%%%%%%%%%%%%%%%%%%%%%%%%%%%%%%%%%%%%%%%%
\subsubsection{Erbium-related complexes and diffusion}
%%%%%%%%%%%%%%%%%%%%%%%%%%%%%%%%%%%%%%%%%%%%%%%%%%%%%%
\label{ssec:Ercomplexes}

Since Er$_{\rm Ca}$ acts as a donor (occurring mainly in a positive charge state), we expect it could form complexes with native defects such as $V_{\rm Ca}$ and ${\rm O}_i$ that act as acceptors (occurring mainly in a negative charge state, see Fig.~\ref{fig:fig_formE_ErMBE}). 
%At point B where the formation energies of Er$_{\rm Ca}$ are the lowest among four representative conditions, $V_{\rm Ca}$ is one of dominant native defects. 
%So, interaction between Er$_{\rm Ca}$ and $V_{\rm Ca}$ would occur in CaWO$_4$. 
Formation energies for these complexes are included in Fig.~\ref{fig:fig_formE_ErMBE}.
For ${\rm Er_{Ca}}-V_{\rm Ca}$, a (0/--) transition level is found at 0.69 eV above the VBM, hence the complex is thermodynamically stable in the negative charge state for most Fermi levels in the band gap.
In the negative charge state of ${\rm Er_{Ca}}-V_{\rm Ca}$, ${\rm Er_{Ca}}$  moves towards the vacancy by 0.06 {\AA} [Fig.~\ref{fig:fig_complex}(c)]. 
We calculate a binding energy $E_{\rm bind} [({\rm Er_{Ca}}-V_{\rm Ca})^-] =   E^f (V_{\rm Ca}^{2-}) + E^f ({\rm Er_{Ca}}^{+}) -E^f [({\rm Er_{Ca}}-V_{\rm Ca})^-]$ = 0.42 eV. 
This modest binding energy indicates that the complex may not form during growth.
However, some degree of complex formation is still expected to occur, given that (1) Er$_{\rm Ca}^+$ is likely to be compensated by acceptors, for which $V_{\rm Ca}$ are the main candidates; (2) $V_{\rm Ca}$ can be mobile during cooldown or annealing (see Sec.~\ref{sec:VCa}); and (3) $V_{\rm Ca}^{2-}$ is Coulombically attracted by the  Er$_{\rm Ca}^+$  donor.
Similarly, when Er is incorporated into CaWO$_4$ by implantation, calcium vacancies that were already in the sample or where created as a result of implant damage could move and bind to Er$_{\rm Ca}$ during subsequent annealing. 
%Chris 051526
Evidence for ${\rm Er_{Ca}}-V_{\rm Ca}$ complexes was indeed found in CaWO$_4$ doped with Er during growth~\cite{Becker2025}.

Formation of a ${\rm Er_{Ca}}-V_{\rm Ca}$ complex will interfere with the proper functioning of Er for quantum applications, given the symmetry lowering and possible change in electronic structure.
For the ${\rm Er_{Ca}}-V_{\rm Ca}$ complexes that form, we can estimate the activation energy for dissociation by adding the migration barrier for the mobile defect species to the binding energy of the complex. 
Since $E_{\rm b}$ of $V_{\rm Ca}^{2-}$ is 1.89 eV, we obtain an activation energy of 1.89 + 0.42 =2.31 eV.
Following the approach outlined in Sec.~\ref{ssec:migration} this implies that an annealing temperature of about 800~K would be needed to dissociate the complex.
% This indicates that once the complex forms, it would be stable at high temperature.  
To remove the vacancy, the anneal should be carried out in a Ca-rich environment; the same goes for the annealing step following Er implantation.

We note that Er itself may diffuse during the annealing. 
%Chris 051326
Er$_{\rm Ca}$ will most likely diffuse assisted by $V_{\rm Ca}$; 
diffusion assisted by Ca$_i$ is less likely since both Er$_{\rm Ca}^+$ and Ca$_i^{2+}$ are positively charged and Coulomb repulsion will prevent the Ca interstitial from approaching the substitutional Er atom.

Given the involvement of native defects, experimental information about self-diffusion of Ca is relevant.
Gupta and Weirick~\cite{GUPTA19672545} reported an activation energy of 55.8 kcal/mol = 2.42 eV for Ca diffusion in CaWO$_4$ single crystals at temperatures between 1388 and 1665~K. 
Our calculated $V_{\rm Ca}$ migration barrier is 1.89 eV, 
but the diffusion activation energy must also account for the formation energy of $V_{\rm Ca}$. To give an activation energy of 2.42 eV,  $E^f (V_{\rm Ca}^{2-})$ should be as low as 0.53~eV.
This is plausible (see Fig.~\ref{fig:fig_formE}) in a Ca-poor environment and given that the crystals in Ref.~\onlinecite{GUPTA19672545} were
doped with with Nd, which (similar to Er) acts as a donor when incorporated on a Ca site and shifts the Fermi level to higher energies, which lowers the formation energy of $V_{\rm Ca}$.
These observations are also consistent with a reported activation energy of Er in LiNbO$_3$ of 2.28--2.44 eV ~\cite{Er_LiNbO3_1996}. 

%Chris 051326
We note that self-diffusion of Ca could also proceed assisted by Ca interstitials, for which we calculated a migration barrier of 0.51~eV (Sec.~\ref{sec:Cai}).
The formation energy of Ca$_i^{2+}$ for plausible Fermi-level positions tends to be higher than that of $V_{\rm Ca}^{2-}$ (see Fig.~\ref{fig:fig_formE}), except under extreme O-poor and Ca-rich conditions; however, values that, combined with $E_{\rm b}$=0.51~eV, would give a diffusion activation energy around 2.4~eV still seem plausible.

Gupta and Weirick~\cite{GUPTA19672545} also reported much lower activation energies for diffusion measured at lower temperatures (980--1388 K).
%Chris 051326
%, along with anomalously small prefactors (10$^{-8}$ cm$^2$/s).
% After excluding the data point at 980 K, they obtained a much larger prefactor of the order of 10$^{-8}$ cm$^2$/s together with an activation energy of 29.5 kcal/mol (1.27 eV). 
This may indicate that a different diffusion process is active; in particular, it was noted that the apparent diffusivities near the surface are smaller than the values obtained from regions deep within the crystal. 

%We note that Er itself may diffuse during the annealing. The activation energy of Er in LiNbO$_3$ was reported to be 2.28--2.44 eV ~\cite{Er_LiNbO3_1996}. In CaWO$_4$, Er$_{\rm Ca}$ will most likely diffuse assisted by $V_{\rm Ca}$. The $V_{\rm Ca}$ migration barrier is 1.89 eV.  But the diffusion activation energy must also account for the formation energy of $V_{\rm Ca}$. To give an activation energy of 2.28--2.44 eV,   $E^f (V_{\rm Ca}^{2-})$ should be as low as 0.39--0.55 eV, which is plausible (see Fig.~\ref{fig:fig_formE}) in a Ca-poor environment and given that doping with Er (which acts as a donor) is likely to shift the Fermi level to higher energies, which lowers the formation energy of $V_{\rm Ca}$

%Note that the $E_{\rm b}$ of Er$^{3+}$ was not considered in the present work, as it is much heavier than Ca, W, and O; the atomic mass of Er is more than 4 times larger than that of Ca (As mentioned above, $E_{\rm b}$ of $V_{\rm Ca}^{2-}$ is quite large).  
%Therefore, the complex formation would be energetically advantageous at the room temperature but complex should be easily dissociated at higher temperature due to small binding energy. 
%The small value of this binding energy indicates that the complex is unlikely to be stable.

%Another plausible complex is negatively charged ${\rm Er_{Ca}}-{\rm O}_i$ because ${\rm O}_i$ would be likely form as much as $V_{\rm Ca}$. 
${\rm O}_i$ is highly mobile (Sec~\ref{sec:Oi}), so a ${\rm Er_{Ca}}-{\rm O}_i$ complex could also form after Er incorporation.
 We find a (+/--) level at 2.36 eV above the VBM. 
In the negative charge state [Fig.~\ref{fig:fig_complex}(d)], the distance between Er and O$_i$ is 2.26 {\AA}, shorter than the Er$_{\rm Ca}$--O bond length and the Ca--O bond lengths in the bulk.
The O$_i$--O$_4$ distance is 2.35 {\AA}, and the distances with the neighboring atoms are 2.02 {\AA} for W$_1$, 2.07 {\AA}  for W$_2$, and 2.90 {\AA}  for Ca$_1$. 
We find a binding energy $E_{\rm bind} [({\rm Er_{Ca}}-{\rm O}_i)^-] =   E^f ({\rm O}_i^{2-}) + E^f ({\rm Er_{Ca}}^{+}) -E^f [({\rm Er_{Ca}}-{\rm O}_i)^-]$ = 0.85 eV.
The resulting activation energy for dissociation would be 0.80 + 0.85 =1.65 eV, thus requiring an annealing temperature of about 600~K to break the complex.
In the positive charge state, the Er--O distances are 2.35 {\AA} for O$_i$ and  2.53 {\AA} for O$_4$, and the O$_i$--O$_4$ distance is 1.46 {\AA}, forming a split-interstitial-like configuration. 
In this charge state, the binding energy is
$E_{\rm bind} [({\rm Er_{Ca}}-{\rm O}_i)^+] =   E^f ({\rm O}_i^{0}) + E^f ({\rm Er_{Ca}}^{+}) -E^f [({\rm Er_{Ca}}-{\rm O}_i)^+]$ = 0.42 eV.
The resulting activation energy for dissociation would the be 0.80 + 0.42 =1.22 eV, implying that annealing at 440~K would already dissociate the complex.

\subsubsection{Discussion}
\label {sec:disc}

We now correlate our results with the effects of annealing observed in optical measurements. 
Ourari {\it et al.}~\cite{2023_Nature_Jeff_ErCWO} reported that a maximum site occupation of ${\rm Er_{Ca}}$ in photoluminescence excitation (PLE) measurement can be achieved by  using an annealing temperature of 300 $^{\circ}$C (= 573 K). 
This indicates that annealing plays a critical role in securing ${\rm Er_{Ca}}$ signals. 
Before annealing, blinking and spectral diffusion were observed, which required cycling to room temperature to be recovered~\cite{Jeff}. Annealing cured this instability. 

%We tried to correlate the occurrence of blinking with the presence of native defects or complexes.
%Since atomic motion is unlikely to occur at the low temperatures of the experiment, an electronic process should be invoked. 
%Blinking could occur if Er$_{\rm Ca}$ changes its charge state; however, we found that only Er$_{\rm Ca}^+$ is stable.
%We thus tentatively conclude that, while native defects may play a role in activation of Er, they cannot explain the observed blinking.
%Chris 051526
We suggest that, after implantation, a sizable fraction of Er atoms reside in an interstitial position.
The incorporation of Er on interstitial sites has been documented in a number of implantation studies in oxides~\cite{Alves1995,Cajzl2022}.
Interstitial Er is still likely to occur in a 3+ charge state (see Sec.~\ref{sssec:Eri}) corresponding to the Er(III) oxidation state that produces the desired emission around 1530~nm.  However, these Er interstitials are, neither atomically nor electronically, as stable as Er$_{\rm Ca}$.
The lower symmetry at the interstitial site will likely lead to line broadening, and the presence of implantation-produced charged native defects will cause electric fields and produce spectral diffusion.
In addition, as shown in Fig.~\ref{fig:fig_formE_ErMBE}, Er$_i$ can occur in charge states other than Er$_i^{3+}$.
An electron could be transferred to Er$_i^{3+}$ from a nearby point defect, either by tunneling or assisted by the optical excitation, and change the charge state to 2+, causing the Er$_i$ to go dark.

Our results in Sec.~\ref{sssec:Eri} indicate that annealing at modest temperatures will allow the Er interstitials to kick into a lattice site, leading to dominantly substitutional Er which, due to the absence of other charge states, is stable in the Er(III) oxidation state.
Combined with the ability of native defects to migrate away (Sec.~\ref{sec:form}), this explains why annealing at 300 $^{\circ}$C results in stable luminescence~\cite{2023_Nature_Jeff_ErCWO}.
Annealing at higher temperatures does not provide benefits, and at a temperature of 800 $^{\circ}$C, luminescence was observed to sharply drop, which can be attributed to Er diffusion, consistent with the discussion in Sec.~\ref{ssec:Ercomplexes}.

\section{Summary}

We have investigated the structural and electronic properties of native point defects 
%such as vacancies and interstitials 
and Er dopants in CaWO$_4$ using hybrid density functional calculations. 
%Oxygen-related defects ($V_{\rm O}$ and O$_i$) and calcium-related defects ($V_{\rm Ca}$ and Ca$_i$) are important defects. 
We find that defects related to the W site are unlikely to form. Oxygen vacancies and calcium vacancies have the lowest energies under intrinsic conditions, but oxygen and calcium interstitials may also form.
%$V_{\rm O}^+$ and $V_{\rm O}^{2+}$ are thermodynamically stable near the mid gap, and 
$V_{\rm O}^+$ has a spin-1/2 ground state, which could contribute to magnetic noise and limit the coherence time of spin qubits. 
The positively charged $V_{\rm O}$ and negatively charged $V_{\rm Ca}$ are likely to bind and form a complex.  
%On the other hand, tungsten-related defects are unlikely to form unless the Fermi level is in the vicinity of the band edges and/or the samples are obtained under extremely oxygen-rich condition, which is unlikely to occur in the wide-band-gap insulator such as CaWO$_4$. 
Calculations of migration barriers allow us to estimate temperatures at which defects would be able to diffuse.  
%Chris 051326
The migration barriers of Ca$_i^{2+}$, $V_{\rm O}^{2+}$ and O$_i^{2-}$ were found to be particularly low (0.51~eV, 0.43~eV and 0.80~eV), indicating high mobility even below room temperature.

%Chris 051326 reordered
We also calculated optical transition levels, allowing us to suggest assignments of experimentally observed absorption or emission peaks.
Our results show that many of the observed signals peaks can be attributed to the oxygen-related defects. 

Erbium impurities prefer to substitute on the Ca site in a positive charge state without significant atomic relaxation. 
${\rm Er_{Ca}}$ can also form complexes with $V_{\rm Ca}$ and O$_i$, which would deactivate the Er. 
%However, the complex with $V_{\rm Ca}$ might be likely with calcium-poor conditions due to the high migration barrier of $V_{\rm Ca}$. 
%Chris 051326
If Er is introduced by implantation, a significant fraction of Er interstitials may be present; our results explain why their related emissions are prone to showing spectral diffusion and blinking.  
Annealing at modest temperatures allows Er atoms to kick into a lattice site and implantation-induced point defects to migrate away, resulting in stable emission from ${\rm Er_{Ca}}$, which is stable only in the Er(III) oxidation state.

%For Er:CaWO$_4$, a maximum site occupation of ${\rm Er_{Ca}}$ in photoluminescence excitation (PLE) measurement, depending on annealing temperature, could be related to light-induced electron transfer between Er and native defects. 

\begin{acknowledgments}

%We gratefully acknowledge fruitful discussions with Prof.~J.~Thompson.
This work was supported by the U.S. Department of Energy, Office of Science, National Quantum Information Science Research Centers, Co-design Center for Quantum Advantage (C2QA) under contract number DE-SC0012704.
M. C. was supported by the Office of Naval Research, Award No.~N00014-22-1-2808 (Vannevar Bush Faculty Fellowship) and by the government of the Republic of Korea (MSIT) and the National Research Foundation of Korea (RS-2024-00466625, RS-2025-16068832, RS-2025-25443942). M.E.T. was supported by the Office of Naval Research through the Naval Research Laboratory's Basic Research Program.
The research used resources of the National Energy Research Scientific Computing Center, a DOE Office of Science User Facility supported by the Office of Science of the U.S. Department of Energy under Contract No. DE-AC02-05CH11231 using NERSC award BES-ERCAP0021021.
%This work used Stampede2 at TACC through allocation DMR070069 from the Advanced Cyberinfrastructure Coordination Ecosystem: Services $\&$ Support (ACCESS) program, which is supported by NSF grants 2138259, 2138286, 2138307, 2137603, and 2138296. 
Use was also made of computational facilities purchased with funds from NSF (No. CNS-1725797) and administered by the Center for Scientific Computing (CSC). The CSC is supported by the California NanoSystems Institute and the Materials Research Science and Engineering Center (No. NSF DMR-2308708) at UC Santa Barbara.
\end{acknowledgments}

%This work was supported by the government of the Republic of Korea (MSIT) and the National Research Foundation of Korea (No. RS-2024-00442710, RS-2024-00466625) 

%\textcolor{red}{**]]}
\bibliography{Choi_CWO}

\end{document}